\documentclass[review, a4paper, 11pt, 3p, sort&compress]{elsarticle}

\usepackage{lineno}

\usepackage[numbers]{natbib}

\usepackage[utf8]{inputenc}
\usepackage[T1]{fontenc}
\usepackage{hyperref}
\usepackage{url}
\usepackage{booktabs}
\usepackage{amsfonts}
\usepackage{nicefrac}
\usepackage{microtype}
\usepackage{graphicx}
\usepackage{doi}
\usepackage{multirow}
\usepackage{optidef}
\usepackage{mathtools}
\usepackage{amsmath}
\usepackage{amssymb}
\usepackage[table]{xcolor}
\usepackage{todonotes}
\presetkeys%
    {todonotes}%
    {inline,backgroundcolor=yellow}{}

\usepackage{soul}
\usepackage[autolanguage]{numprint}
\usepackage[english]{babel}
\usepackage{array}
\usepackage{booktabs}
\usepackage{caption}
\usepackage{tabu}
\usepackage{stackengine}
\usepackage{subfig}
\usepackage[normalem]{ulem}
\usepackage{amsfonts}
\usepackage{cancel}
\usepackage{enumitem}
\usepackage{algorithm}
\usepackage{algpseudocode}

\usepackage[T1]{fontenc}
\usepackage{lmodern}

\usepackage{tabularray}
\usepackage{cellspace}
\usepackage{tablefootnote}
\usepackage{refcount}
\usepackage{bbm}

\usepackage[round-precision=3,round-mode=figures,scientific-notation=true]{siunitx}

\usepackage{longtable}
\usepackage{afterpage}

\usepackage[edges]{forest}
\usetikzlibrary{shadows.blur}
\usetikzlibrary{positioning,chains}

\tikzset{parent/.style={align=center,text width=2cm,fill=green!20,rounded corners=2pt},
    child/.style={align=center,text width=2.8cm,fill=green!50,rounded corners=6pt},
    grandchild/.style={fill=pink!50,text width=2.3cm}
}

\usepackage{breakurl}
\usepackage{etoolbox}
\appto\UrlBreaks{\do\-}

\usepackage{verbatim}

\usepackage{pgfplots}
\pgfplotsset{compat=1.17}
\usetikzlibrary{patterns}

\usepackage{makecell}

\usepackage{doi}

\colorlet{green}{blue}

\hypersetup{pdfauthor=author}

\journal{Electronic Commerce Research and Applications}

\newcommand{\adrev}[1]{\textcolor{black}{#1}}

\newcommand{\quotes}[1]{``#1''}
\newcommand{\noop}[1]{}

\begin{document}

\let\WriteBookmarks\relax
\def\floatpagepagefraction{1}
\def\textpagefraction{.001}

\definecolor{Gray}{gray}{0.9}

\begin{titlepage}

\begin{frontmatter}
\title{Economic Recommender Systems -- A Systematic Review}

\author[1,2]{Alvise De~Biasio}%[type=editor, orcid=0000-0003-0528-6223]
\ead{alvise.debiasio@phd.unipd.it}

\address[1]{Department of Mathematics, University of Padova, Via Trieste 63, Padova, 35131 PD, Italy}

\address[2]{R\&D Department, estilos srl, Via Ca' Marcello 67/D, Venezia, 30172 VE, Italy}

\author[1]{Nicolò Navarin}%[orcid=0000-0002-4108-1754]
\ead{nnavarin@math.unipd.it}

\author[3]{Dietmar Jannach\corref{cor1}}
\ead{dietmar.jannach@aau.at}

\address[3]{Department of AI \& Cybersecurity, University of Klagenfurt, Universitätsstraße 65--67, 9020 Klagenfurt, Austria}
\cortext[cor1]{Corresponding author}

\begin{abstract}
Many of today's online services provide personalized recommendations to their users.
Such recommendations are typically designed to serve certain user needs, e.g., to quickly find relevant content in situations of information overload.
Correspondingly, the academic literature in the field largely focuses on the value of recommender systems for the end user.
In this context, one underlying assumption is that the improved service that is achieved through the recommendations will in turn positively impact the organization's goals, e.g., in the form of higher customer retention or loyalty.
However, in reality, recommender systems can be used to target organizational economic goals \emph{more directly} by incorporating monetary considerations such as price awareness and profitability aspects into the underlying recommendation models.
In this work, we survey the existing literature on what we call \emph{Economic Recommender Systems} based on a systematic review approach that helped us identify 135 relevant papers.
We first categorize existing works along different dimensions and then review the most important technical approaches from the literature. Furthermore, we discuss common methodologies
to evaluate such systems and finally outline the limitations of today's research and future directions.
\end{abstract}

\begin{keyword}
Recommendations
\sep Business Value  \sep Price and Profit \sep Multistakeholder
\sep Survey
\end{keyword}

\end{frontmatter}
\vfill
\end{titlepage}
%\linenumbers
%\maketitle
% =========================================
\section{Introduction}
\label{sec:intro}
% =========================================
\textit{Recommender Systems} (\textit{RSs}) \cite{jannach2010recommender} have become an integral part of many modern online services, for example, on Amazon, Netflix, YouTube, or Spotify.
Typically, the recommendations provided by the system are designed to serve certain user needs.
On the mentioned e-commerce and media streaming sites, for example, these systems support users in navigating large information spaces, thereby helping them discover relevant content that they were previously not aware of.

The academic literature on RSs has traditionally focused on the different types of value that such systems create for \emph{users}, in particular by proposing increasingly sophisticated machine learning models to predict which items are relevant for them in a given situation.
An underlying assumption of this user-centric perspective is that by creating value for consumers through personalized recommendations, providers expect certain benefits for the organization as well, for example, through increased customer engagement, loyalty, and retention \cite{Domingues2013,garcin_offline_2014,Holtz2020Engagement,gomez-uribe_netflix_2016}.

Only during the last few years, researchers increasingly emphasize the fact that in practical applications of RSs, the interests of multiple stakeholders have to be \emph{explicitly} taken into account.
Correspondingly, the underlying systems have to be designed to create value both for consumers, recommendation providers, and maybe even further stakeholders \cite{abdollahpouri_multistakeholder_2020,jannach_recommendations_2016,ricci_value_2022}.

In practice, the business value an RS
creates for providers is measured through various \textit{key performance indicators} (\textit{KPIs}), see  \cite{debiasio_value_2023,jannach_measuring_2019}. Besides the mentioned indirect effects of recommendations on customer engagement and retention, organizations rely on various forms of conversion rates to gauge the effectiveness of a system. In many cases, firms directly assess the impact of recommendations by analyzing the effects on sales numbers \cite{pathak2010empirical, garfinkel_empirical_2007, panniello2016research}.
In particular, the use of side or contextual information \cite{sun2019research, villegas2018characterizing} has proven useful in many circumstances to improve business KPIs \cite{panniello2016research, cooke2002marketing, gorgoglione2011effect, panniello2012incorporating, panniello2014comparing, panniello2009experimental, panniello2009comparing, gorgoglione2009including}.

Therefore, it becomes desirable for companies to incorporate relevant knowledge into the underlying algorithms so that the resulting recommendations can drive these KPIs more directly in the desired direction \cite{panniello2016research, cooke2002marketing, gorgoglione2011effect}.
One important domain-independent approach in this context is to consider purchase-related information in the recommendation models, in particular regarding the profit that results from individual purchase transactions, see \cite{jannach_price_2017}. Moreover, such
algorithms may implement several other theories and mechanisms from the economics and marketing literature,
considering, for example, the role of promotions and discounts, price sensitivity, or consumer utility.

These approaches, which we call \emph{Economic Recommender Systems} (\emph{ECRSs}), are highly relevant in practice.
Unfortunately, the literature on this topic is largely scattered.
With this paper, we provide a systematic review of the field, which should serve researchers and practitioners alike as a starting point to understand the state-of-the-art in the area.
Our systematic literature search surfaced more than one hundred relevant papers, which we categorize into five main dimensions of analysis, see Section~\ref{sec:methodology}.
In the main part of the survey, Section~\ref{sec:technical-approaches}, we then discuss existing ECRSs technical approaches.
Afterward, we analyze existing methodologies
to evaluate such
systems in Section~\ref{sec:experimental-evaluation}.
Finally, we discuss open challenges and future research directions in the field in Section~\ref{sec:open_challenges}.
The paper ends by highlighting some managerial implications of using ECRSs in real-world environments.

% =========================================
\section{Background and Related Work}
\label{sec:background-and-related-work}
% =========================================
In this section, we first provide more background on the business value of recommendations.
We then characterize the concept of ECRSs in more depth. Finally we discuss the relationships of ECRSs
 to neighboring topics in RSs research.
\subsection{Business Value of Recommender Systems}

RSs, as mentioned above, are often designed to serve both user \cite{senecal_influence_2004, bollen_understanding_2010, chen_impact_2004, hinz_impact_2010, haubl_consumer_2000} and organizational purposes \cite{jannach_recommendations_2016, van_capelleveen_recommender_2019, gorgoglione_recommendation_2019, maslowska_role_2022}.
Regarding the organizational purposes,
there are various ways in which an RS can generate value for a business \cite{adomavicius_toward_2005, amatriain_past_2016, gomez-uribe_netflix_2016, amatriain_beyond_2013, schafer_recommender_1999, schafer2001commerce, xiao2007commerce, belluf_case_2012, panniello2014use}, considering economics and
marketing aspects \cite{ting-kai_hwang_optimal_2014, hanafizadeh_insight_2021, oechslein_search_2014, krasnodebski_considering_2016, ren_is_2021}, e.g., by converting website visitors into buyers, increasing cross-selling opportunities or building customer loyalty \cite{schafer2001commerce}.

In the literature, a number of general categories were identified to characterize how RSs may create business value and how the business value can be measured \cite{debiasio_value_2023,jannach_measuring_2019}.
The typical measures and corresponding KPIs include \cite{jannach_measuring_2019}:
\begin{itemize}
    \item The number of \textit{clicks} from recommendations, often measured by the \textit{click-through rate} (\textit{CTR});
    \item The degree of \textit{user adoption} of the system, often measured by the \textit{conversion rate} (\textit{CVR});
    \item The overall \textit{revenue} generated from the \textit{sales} of the firm's products and services;
    \item The possible effects on the \textit{sales distribution} of the items sold, e.g., shifting toward more profitable items;
    \item The overall degree of \textit{user engagement} with the platform, as indicator of \textit{customer satisfaction}.
\end{itemize}

RSs that are designed to optimize one or more of the above business values are generically referred to in the literature as ``value-aware'' \cite{debiasio_value_2023}.
Depending on the business industry (e.g., retail, entertainment, manufacturing) or the revenue model (e.g., transaction-based, advertising, subscription) \cite{resnick_recommender_1997, frew_recommender_2005, herder_need_2019, mehrotra_recommendations_2019, li_recommender_2018, chen_community-based_2009, hoffman_conceptual_2005} the company may want to optimize certain business values rather than others.
For example, in the case where the revenue model is primarily transaction-based (e.g., Walmart), since there is a direct link between purchases and revenue, the company might be interested in shifting the customer behavior towards the purchase of the more profitable items \cite{panniello_impact_2016}.
In contrast, in case the organization's revenue model is
 based on ads (e.g., YouTube), the company may be interested in increasing the number of clicks \cite{davidson2010youtube} as this is directly related to the consumption of ads that providers pay to see their brand advertised.
Finally, a company might also be interested in optimizing user engagement \cite{gomez-uribe_netflix_2016} in the case of subscription-based models (e.g., Netflix) as this positively correlates with retention.

\subsection{Economic Recommender Systems}
\label{sec:ecrs}

While there are various ways in which a value-aware RS can create value for users and providers, and while there are several KPIs that firms might seek to optimize, ultimately, the provision of a recommendation service almost always serves some economic goal of the organization such as profit and growth. However, we note that some forms of value creation are more directly targeting profitability aspects than others. An increase in revenue through recommendations or a shift in the sales distribution toward the most lucrative items is almost directly reflected in a profit improvement \cite{hosanagar_recommended_2008, chen_developing_2008, pei_value-aware_2019}. On the other hand, a growth in user engagement, as in the case of Netflix \cite{gomez-uribe_netflix_2016}, with more customers joining and fewer leaving, is sometimes only indirectly reflected in higher long-term profits for the organization.

In this survey, we focus on the first type of the described recommendation approaches, i.e., a particularly prominent subset of value-aware RSs \cite{debiasio_value_2023} that target economic effects in a more direct way.
Typical examples in this context are: RSs that consider company profit and customer relevance in a balanced way \cite{nemati_devising_2020, concha-carrasco_multi-objective_2023, cai_trustworthy_2019, chen_developing_2008, kompan_exploring_2022}; systems that leverage discounts and pricing algorithms to trigger purchases \cite{jiang_optimization_2012, adelnia_najafabadi_dynamic_2022, zhao_e-commerce_2015, jannach_determining_2017, jiang_redesigning_2015}; or methods that consider
customers' price sensitivity to recommend items more in line with their price preferences \cite{ge_cost-aware_2014, chen_boosting_2017, zheng_price-aware_2020, zheng_incorporating_2021, zhang_price_2022}. We call such systems economic recommender systems, and we informally characterize them as follows:

\begin{quote}
\emph{An Economic Recommender System (ECRS) is an RS that exploits price and profit information and related concepts from marketing and economics to directly optimize an organization's profitability.}
\end{quote}
Later in this work (Section~\ref{sec:methodology}), we identify five key approaches from the literature to build ECRSs, which we divide into customer and organization-oriented ones, depending on the focus of the underlying algorithms.
Customer-oriented approaches in the literature, for instance, integrate purchasing behavior mechanisms (e.g., price sensitivity) into the models to generate more relevant recommendations that will automatically lead to more profit.
Organization-oriented ones, on the other hand,
apply particular
organizational strategies (e.g., profit awareness, promotional pricing)
to optimize profit.

Since most RSs may at least indirectly target some profit-related or growth-related goal, the boundaries between an economic RS and a ``traditional'' one may sometimes appear blurry.
However, a clear distinction can often be made depending on the underlying revenue model \cite{chen_community-based_2009, resnick_recommender_1997, hoffman_conceptual_2005}.
For example, click-through rate maximization may be seen as an indirect method for profit optimization in case it is only about increasing site interactions \cite{guo_we_2021, wu_returning_2017}.
However, it may also be considered as an ECRS method in case there is some revenue associated with each click event (e.g., commissions suppliers pay to marketplaces for each generated impression), as in the case of an advertising revenue model \cite{malthouse_multistakeholder_2019, theocharous_personalized_2015, zhang_smart_2017}.

Concluding our characterization of ECRSs, it is important to note that considering certain types of economic information to an inappropriate extent may also lead to \emph{unintended negative effects} and \emph{behavioral harms} of recommendations \cite{adomavicius_recommender_2022, hazrati_impact_2021, dorner_think_2013}. Specifically, it is vital to ensure that an ECRS does not negatively impact the user's trust  \cite{liao_user_2022} in the organization \cite{hosanagar_recommended_2008, ghanem_balancing_2022, panniello_impact_2016, basu_personalized_2021, panniello2016research}.
Indeed, trust is one of the most important factors driving adoption \cite{university_of_british_columbia_canada_trust_2005, komiak_effects_2006} and purchase intention \cite{nilashi_recommendation_2016, subudhi_influential_2022}.
Recommendations that are irrelevant \cite{chau_examining_2013, fui-hoon_nah_moderating_2015, zhang_personalize_2013, nguyen_more_2022}, manipulative \cite{kocher_new_2019, adomavicius_recommender_2013, adomavicius_recommender_2022, cremonesi_investigating_2012, gretzel_persuasion_2006, xiao_empirical_2018, tsao_influence_2023, DEBIASIO2023110699}, or poorly explainable \cite{wang_effects_2016, cramer_effects_2008, zhang_exploring_2018} because they are too biased towards the profitable items \cite{wang_effects_2019} can harm trust, leading customers to reactance \cite{fitzsimons_reactance_2004, yanping_psychology_2012} or churning.

Besides trust, there are also other possible harms that may emerge in case the recommendation strategy is oriented too strongly toward profit. While algorithms are often
designed to improve sales diversity \cite{kunaver_diversity_2017, adomavicius_improving_2012} or to stimulate the sales or consumption of niche items \cite{matt_differences_2013, yi_recommendation_2022}, they in practice might sometimes nudge users to buy the most popular ones \cite{fleder_recommender_2007, fleder_blockbuster_2009, lee_impact_2014, hosanagar_will_2014, lee_how_2019, fleder_recommender_2010}. Such effects may in turn have profit implications considering that popular items sometimes have lower margins \cite{garfinkel_empirical_2007}.
Finally, competition effects \cite{ghoshal_impact_2015, li_informative_2020, zhou_impact_2022} may also be important to consider, since rewarding higher-margin items could push sellers to increase prices \cite{zhou_competing_2021}, thus impacting customers' willingness-to-pay \cite{adomavicius_effects_2018}, and market demand \cite{zhang_welfare_2021, aridor_recommenders_2022}.

\subsection{Related Areas in RSs Research}
\label{sec:related_surveys}

ECRSs are related to other important research areas, including the following:
\begin{itemize}
    \item \textit{Multi-Stakeholder Recommender Systems} \cite{abdollahpouri_multistakeholder_2020, ricci_multistakeholder_2022}: where the system is designed to
    meet the interests of multiple stakeholders (e.g., consumers, providers, suppliers);
    \item \textit{Multi-Objective Recommender Systems} \cite{zheng_survey_2022, alhijawi_survey_2023}: where the system is designed to optimize several objectives simultaneously (e.g., accuracy, diversity);
    \item \textit{Fair Recommender Systems} \cite{dejdjoo2023fairness, zehlike_fairness_2023, zehlike_fairness_2023_2, patro_fair_2022, pitoura_fairness_2022}: where the system is designed to avoid possible discrimination
    against certain user or item groups.
\end{itemize}

The relationships between ECRSs and these other areas can be characterized as follows. Regarding multi-stakeholder RSs, we note that probably any ECRS in practice does not \emph{exclusively} focus on provider profitability but considers the interests of other stakeholders---in e-commerce, in particular, those of consumers or suppliers as well \cite{cai_trustworthy_2019, chen_developing_2008}.
Such multi-stakeholder considerations
mean that ECRSs in practice are \emph{multi-objective} RSs that consider
different competing objectives,
e.g., profitability vs.~consumer value \cite{concha-carrasco_multi-objective_2023, ghanem_balancing_2022}  or short-term vs.~long-term profits. However, not every multi-stakeholder RS necessarily is an economic one, e.g., considering that an RS
may also be designed to recommend users to other users (e.g., on dating platforms). Likewise, a multi-objective RS could also optimize non-economic goals, e.g., popularity, which may in turn have a direct inverse relationship with profitability under certain circumstances \cite{garfinkel_empirical_2007}.
Finally, in terms of fairness, when building an ECRS there is always the possibility that by designing a system too biased \cite{chen_bias_2023} toward profitable items \cite{chen_developing_2008, mu-chen_chen_hprs_2007}, the organization might risk being perceived as unfair by consumers. However, there are various other application areas of fair RSs, which are not related to economic aspects or firm profitability, e.g., when the recommender system is designed to avoid discrimination of underrepresented groups in the recommendations.

There are already several surveys on RSs that offer generic introductions \cite{ko_survey_2022}, or focus on certain algorithmic aspects such as deep learning \citep{zhang2019deep, dau_recommendation_2020} or context awareness \cite{sun2019research, villegas2018characterizing}.
In addition, various surveys have been published in the above mentioned areas of multi-stakeholder \cite{abdollahpouri_multistakeholder_2020, ricci_multistakeholder_2022} and multi-objective \cite{zheng_survey_2022, alhijawi_survey_2023} RSs, and on related topics such as fairness \cite{zehlike_fairness_2023, zehlike_fairness_2023_2, patro_fair_2022, pitoura_fairness_2022}, diversity \cite{kunaver_diversity_2017}, trust \cite{dong_survey_2022}, and explainability \cite{zhang_explainable_2020, tintarev_survey_2007}. We refer the readers to these important works for in-depth coverage of the respective topics. The present survey has certain affinities with previous reviews on value-aware \cite{debiasio_value_2023} and price- and profit-aware \cite{jannach_price_2017} RSs.
It however differs from these previous works in various ways.
First, our study is the first systematic review of ECRSs based on what in the literature are denoted as PRISMA guidelines \cite{page_prisma_2021}, i.e., a systematic article review process, known throughout the literature for its high reliability,
which aims to identify all available research relevant to a set of research questions. Moreover, existing value-aware RSs research \cite{debiasio_value_2023} investigated how to generically optimize business value through RSs, whereas our research on ECRSs is focused on direct optimization of profitability.
This subset of value-aware RSs is of particular interest to companies because, as we noted in Section~\ref{sec:ecrs}, ultimately, the provision of a recommendation service almost always serves some economic goal of the organization such as profit and growth.
In addition, the present research on ECRSs also embraces customer-oriented approaches (e.g., price-sensitive recommendations) that had not been studied by previous research on value-aware RSs.
Furthermore, previous work on price- and profit-aware RSs \cite{jannach_price_2017} also focused on profitability optimization.
However, this earlier research did not cover a number of important approaches that were identified in this survey (e.g., economic utility modeling methods).
Finally, our present review also discusses methodological questions (e.g., performance evaluation methods) that were not addressed in previous works.

\section{Methodology}
\label{sec:methodology}

The present study follows a systematic review process based on \textit{Preferred Reporting Items for Systematic Reviews and Meta-Analyses} (\textit{PRISMA}) \cite{page_prisma_2021} guidelines.
The PRISMA article selection process is recognized throughout the scientific community as a rigorous and reliable methodology.
The process aims to identify, evaluate, and interpret all available research relevant to a particular research question, topic area, or phenomenon of interest while ensuring high reproducibility of results.
In the following, we report:
the dimensions of analysis
considered in the study,
the underlying research questions,
the eligibility criteria for article inclusion,
the search queries used, the overall article analysis and selection process,
and the possible limitations of the survey.

\subsection{Decomposing Economic Recommender Systems into Different Dimensions of Analysis}
\label{sec:decomposition}

\begin{figure}
\begin{center}
%\begin{forest}
%  basic/.style = {draw, thin, drop shadow, font=\sffamily, align=center},
%  upupper style/.style = {basic, edge+={-, line width=.4pt}, fill=black!24},
%  upper style/.style = {basic, edge+={-, line width=.4pt}, fill=black!12},
%  lower style/.style = {basic, edge+={-, line width=.4pt}, fill=black!2},
%  where level<=2{%
%    upper style,
%    edge path'={
%      (!u.parent anchor) -- +(0,-7pt) -| (.child anchor)
%    },
%  }{%
%    lower style,
%  },
%  where level<=1{%
%    parent anchor=children,
%    child anchor=parent,
%    if={isodd(n_children())}{%
%      calign=child edge,
%      calign primary child/.process={
%        O+nw+n{n children}{(#1+1)/2}
%      },
%    }{%
%      calign=edge midpoint,
%    },
%  }{
%    folder,
%    grow'=0,
%  },
%  [Economic Recommender Systems\\Dimensions of Analysis, upupper style
%    [Customer Oriented\\ Approaches, folder, grow'=0, for children={lower style},
%      before drawing tree={
%        tempdima/.option=!r2.max y,
%        tempdima-/.option=max y,
%        for tree={
%          y+/.register=tempdima,
%        },
%      }
%      [Price Sensitivity]
%      [Economic Utility\\ Modeling]
%    ]
%    [Organization Oriented\\ Approaches, folder, grow'=0, for children={lower style},
%      before drawing tree={
%        tempdima/.option=!r2.max y,
%        tempdima-/.option=max y,
%        for tree={
%          y+/.register=tempdima,
%        },
%      }
%      [Profit Awareness]
%      [Promotional]
%      [Long-Term Value\\ Sustainability]
%    ]
%  ]
%\end{forest}
\includegraphics[height=6.8cm,width=8cm]{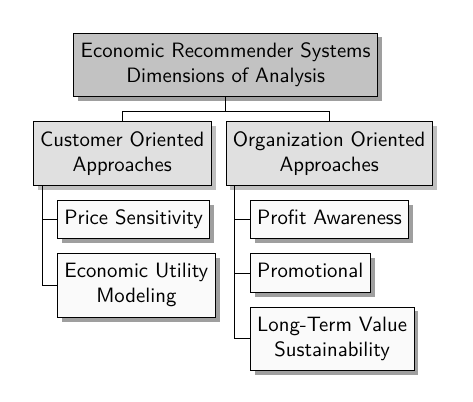}
\end{center}
\vspace{-0.3cm}
\caption{Economic recommender systems dimensions of analysis.}
\label{fig:fig00}
\end{figure}

ECRSs can be characterized by several interrelated topics.
To identify relevant articles, we therefore followed an inductive process starting from two related surveys \cite{jannach_price_2017, debiasio_value_2023}, decomposing ECRSs into different \textit{dimensions of analysis} (\textit{DAs}).
Specifically, we identified five types of approaches by extending a taxonomy originally presented in a well-known paper in the field, i.e., \cite{jannach_price_2017}.
Four of the DAs that we introduce below, i.e., price sensitivity, profit awareness, promotional and long-term value sustainability, have naturally emerged from the analysis of this article.
The fifth dimension, i.e., economic utility modeling, has been identified after a more in-depth analysis of some of the works cited in the article, e.g., \cite{wang_utilizing_2011}.
In particular, we observed that articles such as the one mentioned here exploited techniques and concepts from the economics literature that are different from those that are used in the articles related to the other DAs. Since we also noticed that recently various other articles that exploited similar techniques were published, we decided to group them into a separate DA.
As we will also report later in more detail in Section~\ref{sec:technical-approaches}, given that we identified a substantial number of works for each DA, we are confident that our categorization scheme properly reflects the various types of activities in this research area.

As Figure~\ref{fig:fig00} shows, the five types of approaches
can be divided into customer and organization-oriented ones, depending on their main focus.
Customer-oriented approaches aim to integrate RSs models with purchasing behavior mechanisms to generate more relevant recommendations that could in turn lead to more value for the firm.
Instead, organization-oriented ones make use of specific organizational strategies to directly or indirectly optimize business KPIs.
Intuitively, the main difference between the two types of approaches is that customer-oriented ones tend to consider the problem from the customer's perspective,
while organization-oriented ones tend consider the problem from the organization's perspective.
In particular, customer-oriented approaches are primarily designed to help users find items that are more in line with their needs.
As a direct result of pursuing this goal, they also help the company make more profit through an expected increase in sales volume.
Instead, organization-oriented approaches are mainly designed to push users toward what help the company improving its business KPIs.
Hence, algorithms may not recommend what is best for users, as they are designed to recommend what is best for the company in terms of profitability gains.
Below, we explain the rationale behind each approach.
\begin{itemize}
    \item \textbf{DA1}: \textbf{Price Sensitivity} approaches aim to explicitly consider customers' price preferences in the recommendation process.
    In fact, price is one of the variables that most strongly influence customers' buying behavior \cite{lichtenstein_price_1993, ali_marketing_2021}.
    For example, customers are often willing to pay more for certain types of items based on presumed greater utility, better aesthetics, brand prestige, supplier reliability, or a combination of various factors \cite{jannach_determining_2017}.
    By considering customers' price sensitivity in the algorithms \cite{jannach_price_2017}, more accurate and relevant recommendations could directly increase the probability of purchase and thus lead to higher sales revenue for the organization.

    \item \textbf{DA2}: \textbf{Economic Utility Modeling} approaches aim to explicitly consider the utility of recommendations for the customer in accordance with an economic perspective.
    There are many utilitarian dynamics \cite{scott2000rational} related to the particular type of purchased products \cite{wang_utilizing_2011, ge_maximizing_2019}.
    For example, if a customer has just purchased a computer or a smartphone, it is very likely that he or she will not purchase the same or a similar product again within a short time.
    Conversely, there are other products, such as dog food or diapers, for which he or she is very likely to continue buying repeatedly for an extended period of time.
    Generating more relevant recommendations by considering the customer's utilitarian behavior could increase conversion rates and generate more profits for the firm.

    \item \textbf{DA3}: \textbf{Profit Awareness} approaches aim to directly incorporate profit information into the recommendation models.
    In fact, profit (i.e., sales revenue minus costs) is one of the most important business KPIs for a successful enterprise \cite{teece_internal_1981}.
    Depending on the particular level of this indicator, a company may or may not invest in research and development to grow the business, attract investors to finance its operations, obtain possible financing from banks, and many other issues of strategic interest to entrepreneurs and managers \cite{geroski_profitability_1993}.
    Overall, generating more profitable recommendations by explicitly considering profit information could directly optimize the organization's economic goals.

    \item \textbf{DA4}: \textbf{Promotional}
    approaches generate recommendations while strategically setting the prices of certain products or focusing the customer's attention on certain brands or promotions.
    For example, the company
    can offer certain products at a discounted price (individually or in bundles) to incentivize impulsive buying behaviors \cite{morgan_research_2019, goi_review_2009}.
    Similarly, the firm can make customers aware of certain products that they would be unlikely to discover on their own and indirectly trigger a possible purchase in the future \cite{kotler1990marketing}.
    Both approaches can be integrated into the recommendation process to optimize profit.

    \item \textbf{DA5}: \textbf{Long-Term Value Sustainability} approaches aim to generate recommendations considering a long-term economic perspective.
    In fact, long-term sustainable business growth is one of the most important aspects for a company \cite{lumpkin_long-term_2010, rauch_effects_2005, ortiz-de-mandojana_long-term_2016}.
    For example, a company may be interested in making customers progressively purchase more and more products and services over time to increase their customer lifetime value.
    Generating recommendations by considering such long-term economic goals of the company thus has the potential to stimulate business growth in a sustainable way over time.

\end{itemize}

\begin{table*}[t]
\footnotesize
	\begin{center}
    \begin{tabular*}{\linewidth}{m{0.5cm} m{2.cm} m{6.cm} c c c c c }
    \toprule
    ID & Dimension & Search Query  & \href{https://www.scopus.com/search/form.uri}{Scopus} & \href{https://ieeexplore.ieee.org/Xplore/home.jsp}{IEEE} & \href{https://link.springer.com/}{Springer} & \href{https://dl.acm.org/}{ACM} & Total \\
    \midrule

    \textbf{DA1} & Price Sensitivity &  ((``recommender system'') AND (``price preference'' OR ``price sensitivity'' OR ``price elasticity'' OR ``willingness to pay'' OR ``price-aware'')) & 670 & 2 & 469 & 57 & \textbf{1198} \\
    \textbf{DA2} & Economic Utility \newline Modeling & ((``recommender system'') AND (``economic'') AND (``utility theory'')) & 104 & 0 & 188 & 17 & \textbf{309} \\
    \textbf{DA3} & Profit Awareness & ((``recommender system'') AND (``multi-stakeholder'' OR ``profit-aware'' OR ``value-aware'')) & 351 & 5 & 290 & 63 & \textbf{709}\\
    \textbf{DA4} & Promotional
    & ((``recommender system'') AND (``dynamic pricing'' OR ``price personalization'' OR ``product bundling'')) & 483 & 0 & 423 & 29 & \textbf{935} \\
    \textbf{DA5} & Long-Term Value \newline Sustainability & ((``recommender system'') AND (``customer lifetime value'' OR ``RFM'' OR ``cumulative profit'' OR ``long-term value'')) & 619 & 4 & 450 & 35 & \textbf{1108} \\

    \bottomrule
    \end{tabular*}
    \end{center}
    \vspace{-0.3cm}
    \caption{Search queries and results divided by online database of the different dimensions of analysis on which this article focuses. Queries were run on \textit{May 12, 2023 looking for all documents published since \textit{January 1, 2000}.}}
	\label{tab:search_queries}
\end{table*}

\begin{figure*}[t]
\footnotesize
	\begin{center}
    	\includegraphics[width=\linewidth]{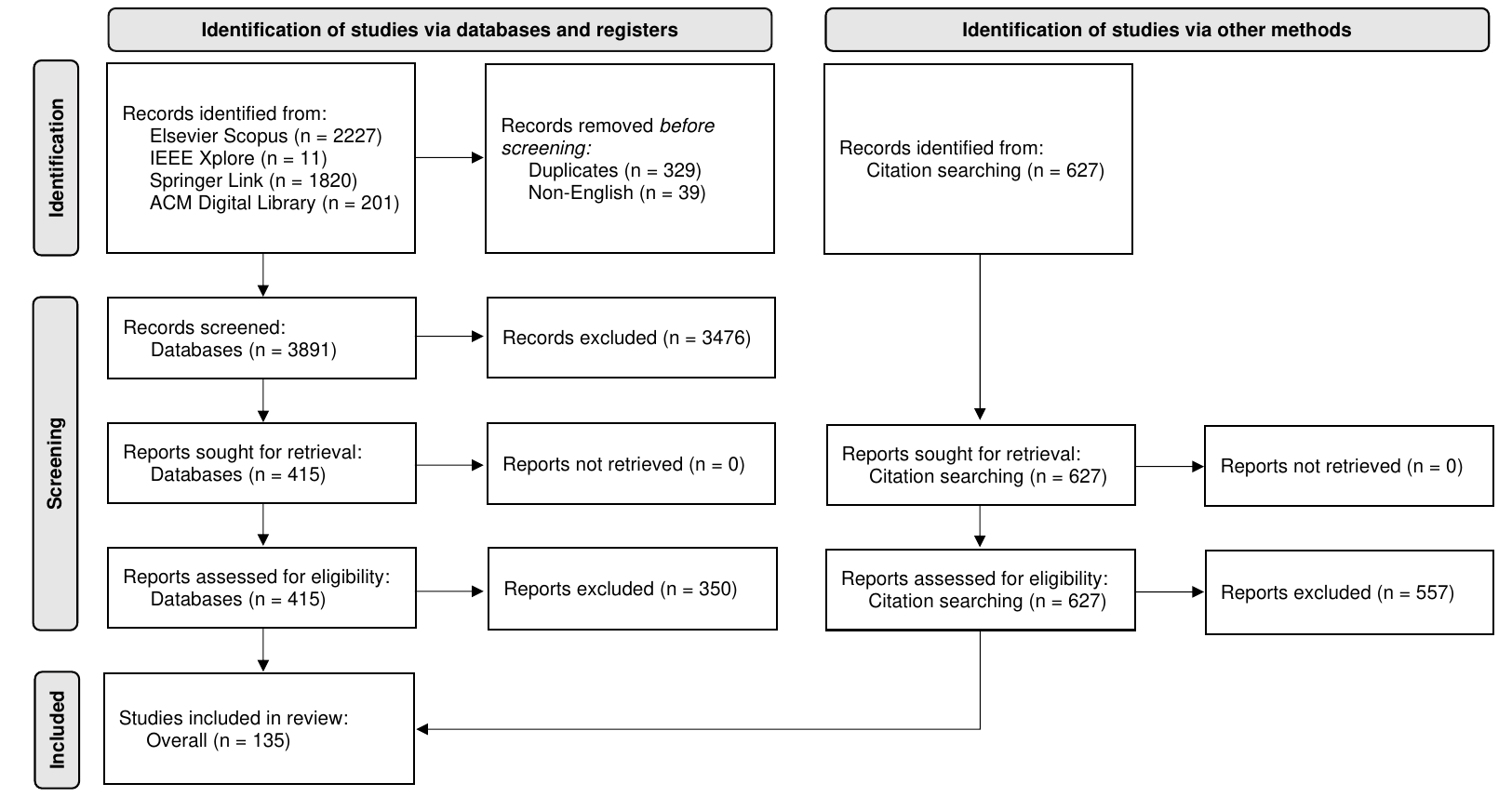}
    \end{center}
    \vspace{-0.3cm}
	\caption{PRISMA flow diagram.}
	\label{fig:fig0}
    \vspace{-0.5cm}
\end{figure*}

\begin{table}[tp]
\tiny
	\begin{center}
    \begin{tabular}{m{6.5cm} c m{6.5cm} c }
    \toprule
    Top-10 Conferences & \#Ref. & Top-10 Journals & \#Ref. \\
    \midrule
ACM Conference on Recommender Systems (RecSys) & 11 & Expert Systems with Applications & 4  \\
ACM Conference on Research and Development in Information Retrieval (SIGIR) & 7 & IEEE Transactions on Knowledge and Data Engineering & 4  \\
International Conference on World Wide Web (WWW) & 7 & Decision Support Systems & 4  \\
ACM International Conference on Knowledge Discovery and Data Mining (SIGKDD) & 7 & Electronic Commerce Research and Applications & 2  \\
ACM Conference on Information \& Knowledge Management (CIKM) & 6 & Knowledge-Based Systems & 2  \\
International Conference on Data Mining (ICDM) & 4 & ACM Transactions on Information Systems & 2  \\
ACM Conference on Web Search and Data Mining (WSDM) & 2 & Information Sciences & 2  \\
IEEE Conference on Data Engineering (ICDE) & 2 & Electronic Markets & 1  \\
International Conference on Information Systems (ICIS) & 2 & Future Generation Computer Systems & 1  \\
European Conference on Information Systems (ECIS) & 2 & IEEE Intelligent Systems & 1  \\     \bottomrule
    \end{tabular}
    \end{center}
    \vspace{-0.3cm}
    \caption{Top-10 conferences and journals ranked by number of published articles from the surveyed literature.}
    \vspace{-0.1cm}
	\label{tab:top_venues}
\end{table}

\subsection{Research Questions}

Having identified these DAs, the goal of our work is to review the state-of-the-art of current ECRSs research.
More specifically, the present survey aims to answer the following \textit{research questions} (\textit{RQs}):
\begin{itemize}
    \item \textbf{RQ1}: What technical approaches are used to build ECRSs?
    \item \textbf{RQ2}: What evaluation methods are used to assess the performance of an ECRS?
    \item \textbf{RQ3}: What are the main challenges and future research directions in the area of ECRSs?
\end{itemize}

\subsection{Search Queries}

As mandated by the PRISMA guidelines, our survey aims to answer previous RQs by systematically querying online libraries.
In particular, taking inspiration from two related systematic review in the RSs field, i.e., \cite{debiasio_value_2023, nunes_systematic_2017}, we queried Elsevier Scopus, IEEE Xplore, Springer Link, and ACM Digital Library to identify relevant articles.
Those online databases are known throughout the literature to contain all the major research works in computer science. For each of these databases, we created a \textit{search query} for  the previous DAs by analyzing the most recurring key terms identified in a series of specialized articles extracted from the literature of two related surveys \cite{jannach_price_2017, debiasio_value_2023}.
In Table~\ref{tab:search_queries}, we report the used search queries and the number of identified documents.

\subsection{Eligibility Criteria}

To be included in the review, articles must pass a rigorous analysis process.
Specifically, articles must meet the following \textit{eligibility criteria} (\textit{EC}):
\begin{itemize}
    \item \textbf{EC1}: Articles must focus on research questions related to one of the dimension of analysis of ECRSs.
    \item \textbf{EC2}: Articles must explicitly mention the business KPIs included in the search queries.
    \item \textbf{EC3}: Articles must be unique, written in English, and the full content must be accessible to the authors.
    \item \textbf{EC4}: Articles must be peer-reviewed by either scientific journals or conferences.
    \item \textbf{EC5}: Graduate theses and doctoral dissertations are not included.
\end{itemize}

\subsection{Article Selection Process}

\begin{figure*}[t]
\footnotesize
    \begin{center}

    \includegraphics[height=7cm]{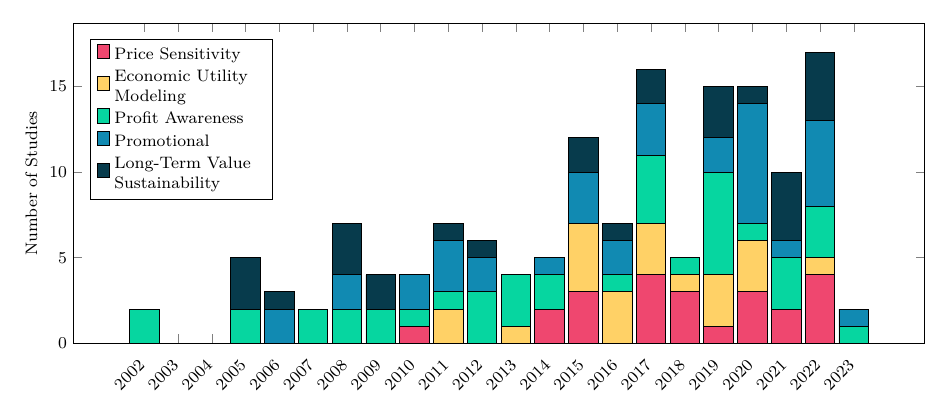}
    \end{center}
    \vspace{-0.3cm}
    \caption{Distribution of surveyed papers per year divided by dimension of analysis.}
    \label{fig:analsys_identified_documents}%
\end{figure*}

As shown in the PRISMA flow diagram in Figure~\ref{fig:fig0}, we followed a multi-stage process to identify all the relevant resources included in this review.
In the first identification phase $2227$ articles from Elsevier Scopus, $11$ articles from IEEE Xplore, $1820$ articles from Springer Link, and $201$ articles from ACM Digital Library were identified for subsequent analyses.
In this phase, $329$ duplicated records and $39$ non-English articles were identified and removed.
In the second screening phase, the titles and abstracts of the remaining $3891$ articles were analyzed, and $3476$ records were removed because the covered topics were not relevant to the present review.
In this phase, $415$ articles were then sought for retrieval and assessed for eligibility, excluding $350$ articles after full text reading.
From this subset of $65$ eligible articles, an additional $627$ articles were identified by searching for references in their bibliographies.
These articles were then assessed for eligibility,
removing $557$ records after reading the full text.
At the end of this overall process, $135$ studies were included in the review.
Of these $135$ articles included in the review, $58.52 \%$ were published in international conferences, $38.52 \%$ in scientific journals and the remaining $3.01 \%$ in book chapters.
As can be seen from Table~\ref{tab:top_venues}, where we report the top-$10$ conferences and journals by number of published articles, many researchers published their work in well-known venues in the field of computer science and particularly in the realm of recommender systems.
In Figure~\ref{fig:analsys_identified_documents}, we show some statistics of the references obtained at the end of the analysis process by reporting the distribution by year of surveyed papers divided by DA.
As can be seen from the figure, there is a growing interest in the literature for all ECRSs dimensions.

\subsection{Study Limitations}

The possible study limitations (SL) are the following:
\begin{itemize}
    \item \textbf{SL1}: Articles were primarily selected from Elsevier Scopus, IEEE Xplore, Springer Link, ACM Digital Library, and from reference searches in the bibliographies of articles that passed the screening stage. Additional online libraries may be considered in future research.
    \item \textbf{SL2}: The study does not cover preprints, non-English articles, non-accessible articles, graduate theses, doctoral dissertations, industry products, and demos.
    \item \textbf{SL3}: Other dimensions of analysis of ECRSs beyond those identified in Section~\ref{sec:decomposition} are left for possible future extensions of this work.
    \item \textbf{SL4}: The study focuses mainly on algorithmic aspects of ECRSs and leaves the in-depth analysis of their managerial implications for the future.
    \item \textbf{SL5}: Although the PRISMA review process can be considered very reliable overall, there is still a small degree of discretionality in the definition of research questions, the selection of keywords to be included in search queries, and the application of eligibility criteria.
\end{itemize}

\begin{table*}[tp]
\newcolumntype{C}[1]{>{\centering\arraybackslash}m{#1}}
\footnotesize
	\begin{center}
    \begin{tabular*}{\linewidth}{C{1.3cm} | m{2.5cm}m{2.5cm}m{2.5cm}m{2.5cm}m{2.5cm}}
    \toprule
    Approach & Price Sensitivity & Economic Utility \newline Modeling & Profit Awareness & Promotional
    & Long-Term Value \newline Sustainability \\
    \midrule
    \rotatebox{90}{In-Processing} & \cite{umberto2015developing, zhang_price_2022, wu_cheaper_2022, maragheh_prospect-net_2022, zheng_incorporating_2021, chen_neural_2021, zheng_price-aware_2020, wang_personalized_2020, kearfott_personalized_2019, greenstein-messica_personal_2018, greenstein-messica_personal-discount_2017, chen_boosting_2017, sato_discount_2015, ge_cost-aware_2014, chen_does_2014} & \cite{maragheh_prospect-net_2022, xu_e-commerce_2020, wang_personalized_2020, ge_maximizing_2019, zhao_multi-product_2017, zhang_economic_2017, zhang_economic_2016, deng_utility-based_2015, wang_utilizing_2011} & \cite{li2017utility, concha-carrasco_multi-objective_2023, dookeram_recommender_2022, nemati_devising_2020, li_revman_2021, ma_placement-and-profit-aware_2019, lin_pareto-efficient_2019, cai_trustworthy_2019, wu_turning_2018, qu_cost-effective_2014, huang_incorporating_2013, goyal_recmax_2012, piton_capre_2011, akoglu_valuepick_2010, kobti_profit-based_2007, wang_mining_2005, brand_random_2005, goos_profit_2002, wang_item_2002}  & \cite{chang_bundle_2023, sun_revisiting_2022, avny_brosh_bruce_2022, agarwal_improving_2022, ghoshal_recommendations_2021, wang_personalized_2020, deng_personalized_2020, chang_bundle_2020, kouki_product_2019, bai_personalized_2019, liu_modeling_2017, ge_effects_2017, sato_discount_2015, massoud_conceptual_2012, wu_enhanced_2011, kamishima_personalized_2011, kowatsch_knowledge-based_2008, garfinkel_shopbot_2008, garfinkel_design_2006} & \cite{iwata_recommendation_2008, iwata_recommendation_2006, chbeir_click_2022, he_propn_2022, zhan_towards_2021, ji_reinforcement_2021, guo_we_2021, zhao_maximizing_2020, zou_reinforcement_2019, pei_value-aware_2019, wu_returning_2017, ju_reinforcement_2017, theocharous_personalized_2015} \\
    \midrule
    \rotatebox{90}{Post-Processing} & \cite{kompan_exploring_2022, cavenaghi_online_2022, wadhwa_personalizing_2020, zhang_identifying_2018, guo_recommend_2018, yang_real_2017, jannach_determining_2017, backhaus_incorporating_2010} & \cite{dai_u-rank_2020, zheng_utility-based_2019, ren_two-sided_2019, zheng_utility-based_2018, yang_real_2017, zhao_e-commerce_2015, adamopoulos_unexpectedness_2015} & \cite{ghanem_balancing_2022, seymen_making_2022, kompan_exploring_2022, ren_is_2021, huang_cost-based_2021, malthouse_multistakeholder_2019, malthouse_algorithm_2019, louca_joint_2019, zhang_smart_2017, yang_real_2017, azaria_movie_2013, wang_strategy-oriented_2012, wang_mathematical_2009, das_maximizing_2009, chen_developing_2008, mu-chen_chen_hprs_2007, lu_show_2014} & \cite{seymen_making_2022, adelnia_najafabadi_dynamic_2022, kini_revenue_2020, guo_maximizing_2020, ettl_data-driven_2020, jannach_determining_2017, demirezen_optimization_2016, beladev_recommender_2016, zhao_e-commerce_2015, jiang_redesigning_2015, zhu_bundle_2014, jiang_optimization_2012, jiang_optimizing_2011, backhaus_incorporating_2010, backhaus_enabling_2010} & \cite{hosein_recommendations_2019, hosanagar_recommended_2008, panniello_impact_2016, basu_personalized_2021} \\
    \bottomrule
    \end{tabular*}
    \end{center}
    \vspace{-0.3cm}
    \caption{ECRSs studies divided by dimension of analysis and algorithmic approach.}
    \vspace{-0.5cm}
	\label{tab:all_approaches}
\end{table*}

\section{Technical Approaches}
\label{sec:technical-approaches}

In this section, we discuss\footnote{The organization of the discussion of the various algorithmic approaches in the next sections varies from section to section.
While this might appear
inconsistent at first glance, this organization is a deliberate choice, because we believe that the current organization of the content is more useful for readers to understand the various concepts.
For example, in Section~\ref{sec:utilitarian}, since the methods presented are based on specific economic theories, we put special emphasis on the underlying utility theories.
In contrast, since in Section~\ref{sec:long_term} long-term value methods based on reinforcement learning represent a very important part of the literature, we devote a separate section to them.} the underlying algorithmic approaches to each of the ECRSs dimensions of analysis introduced in Section~\ref{sec:decomposition}, i.e., price sensitivity, profit awareness, promotional,
long-term value sustainability, and economic utility modeling.

In Table~\ref{tab:all_approaches} we report the studies identified by the present survey that propose technical approaches, categorized by DA and algorithmic method.
The approaches can be divided into in- and post-processing\footnote{Pre-processing methods may also exist in industry, e.g., when a recommendation provider wants to  rule out certain unprofitable items. Our literature search, however, did not surface such approaches.} methods, depending on the time the economic value optimization occurs.
In-processing approaches aim to incorporate economic aspects directly into the models, either by extending the objective function of known algorithms (e.g., by introducing new variables or regularizers) or by developing entirely new algorithms.
The underlying algorithms may be based, for example, on supervised or reinforcement learning paradigms.
Post-processing approaches, on the other hand, can be mounted on top of any recommender and aim to transform the recommendations generated from the baselines by applying specific heuristic economic criteria.
These may incorporate economic value by simply re-ranking the output of the original algorithm or by exploiting additional models.

Analyzing the distribution of the studies in Table~\ref{tab:all_approaches}, we can make some observations.
In particular, it can be noted that there are several relevant works for all the
DAs.
In addition, in-processing and post-processing methods are equally used across all dimensions.
This implies that the research field is broad and that there are various important lines of active research.
Overall, given that there is a substantial number of works in each DA, we are confident that our categorization scheme properly reflects the various types of activities in this research area.

\begin{table}[tp]
\footnotesize
	\begin{center}
    \begin{tabular}{c l}
    \toprule
    Notation & Definition \\
    \midrule

    $u$ & user \\
    $i$ & item \\
    $p_i$ & item's price \\
    $c_i$ & item's cost \\
    $v_i = p_i - c_i$ & item's profit \\
    $m$ & number of overall users \\
    $n$ & number of overall items \\
    $k$ & number of items to recommend \\
    $\mathcal{U} = \{ u_1, \dots, u_m \}$ & set of users \\
    $\mathcal{I} = \{ i_1, \dots, i_n \}$ & set of items \\
    $\mathbf{X} \in \{0,1\}^{m \times n}$ & user-item interaction matrix \\
    $x_{u,i} \in \{0,1\}$  & user-item feedback \\
    $\mathbf{\Theta}$ & set of model parameters \\
    $\mathbf{\hat{X}}(\mathbf{\Theta})$ & scoring function \\
    $\hat{x}_{u,i}(\mathbf{\Theta}) \in [0, 1]$  & user-item predicted interest \\
    $\mathcal{Y}_{u,k}$ & recommendations list \\
    $\mathcal{T}(\mathcal{Y}_{u,k})$ & utility function \\
    $\rho_{u,i}$ & user-item interaction utility \\

    \bottomrule
    \end{tabular}
    \end{center}
    \vspace{-0.3cm}
    \caption{Main notation.}
    \vspace{-0.5cm}
	\label{tab:notation}
\end{table}

\paragraph{Notation}
In the following, we introduce the main notation used in the paper, see Table~\ref{tab:notation}.
Formally, the vast majority of the approaches we discuss in this survey refer to the \textit{top-k recommendation problem} \cite{ricci_item_2022}, i.e., the problem of determining the best $k$ items to recommend to each user.
All algorithms designed to address this particular problem consider a set $\mathcal{U} = \{ u_1, \dots, u_m \}$ of $m$ users, a set $\mathcal{I} = \{ i_1, \dots, i_n \}$ of $n$ items, and a user-item interaction matrix $\mathbf{X}$, where each entry $x_{u,i}$ represents the feedback from a user $u$ towards an item $i$.
With very few exceptions, the feedback considered is almost always implicit (i.e., $x_{u,i} \in \{0,1\}$).
This indicates a positive or missing interaction, depending if the user interacted with the item or not (e.g., purchased it).
Generally, it is assumed that purchased items are those that are relevant (and maybe satisfactory) for consumers.

Algorithms are often designed \cite{ricci_item_2022} to learn a \textit{scoring function} $\mathbf{\hat{X}}(\mathbf{\Theta}): \mathbf{X} \xrightarrow{} \{ \hat{x}(\mathbf{\Theta}) \in \mathbb{R}: 0 \le \hat{x}(\mathbf{\Theta}) \le 1 \}^{m \times n}$ to predict the missing entries of $\mathbf{X}$.
The scoring function is parameterized by a set $\mathbf{\Theta} \in \mathbb{R}^o$ of model parameters\footnote{For some algorithms, such as User-Based Collaborative Filtering based on Nearest-Neighbor techniques \cite{ricci_trust_2022}, we assume $\mathbf{\Theta} = \emptyset$ since there are no model parameters.} - where $o$ is the number of parameters.
Hence, $\hat{x}_{u,i}(\mathbf{\Theta})$ represents the expected interest of the user toward an item he or she has never interacted with.

In the general top-$k$ recommendation problem \cite{adomavicius_toward_2005}, an ordered list $\mathcal{Y}_{u,k}$ of $k$ items to be recommended to each user $u$ is determined optimizing a specific \textit{utility function} $\mathcal{T}(\mathcal{Y}_{u,k}): \mathcal{Y}_{u,k} \xrightarrow{} \mathbb{R}$.
More formally:
\begin{equation}
    \underset{\mathcal{Y}_{u,k}}{\mathrm{argmax}} \quad \mathcal{T}(\mathcal{Y}_{u,k})
\end{equation}
The utility function can be implemented in arbitrary ways (e.g., including relevance, profitability, and other aspects).

Given $\rho_{u,i}$ as the
utility of the user-item interaction, the vast majority of the studies in the RSs literature operationalize the utility function as:
\begin{equation}
    \begin{aligned}
    & \mathcal{T}(\mathcal{Y}_{u,k}) = \sum_{i \in \mathcal{Y}_{u,k}} \rho_{u,i}
    \end{aligned}
\end{equation}
optimizing directly:
\begin{equation}
    \underset{\mathcal{Y}_{u,k}}{\mathrm{argmax}} \sum_{i \in \mathcal{Y}_{u,k}} \hat{x}_{u,i}(\mathbf{\Theta})
    \label{eq:rerank_base}
\end{equation}
and thus considering the
utility of the potential interaction as the expected interest, i.e., $\rho_{u,i} = \hat{x}_{u,i}(\mathbf{\Theta})$.

However, although this user-focused utilitarian conception is currently the most widely used one in the literature, a recommendation provider may have different goals.
In the context of ECRSs, instead, the utility functions may be operationalized considering the item's price $p_i$, and profit $v_i = p_i - c_i$, where $c_i$ is the item's cost.
For example, algorithms belonging to the profit-aware subdomain that we discuss in Section~\ref{sec:profit_aware} are often developed to find the most profitable, yet relevant, items for the company, and these may clearly differ from the \emph{most} relevant ones.

\subsection{Price-Sensitivity Methods}
\label{sec:price_sens}

Price is one of the variables that most influence customers' buying behavior \cite{lichtenstein_price_1993, ali_marketing_2021}.
Accordingly, many studies in the literature \cite{ge_cost-aware_2014, chen_boosting_2017, greenstein-messica_personal_2018, zheng_incorporating_2021, kompan_exploring_2022} propose algorithms to explicitly consider customers' price sensitivity,
as more accurate and relevant recommendations (i.e., in terms of being in the right price range) could directly increase the probability of purchase and thus lead to higher sales revenue for the organization.
Below, we give some insights on how these methods work by discussing a set of selected articles.

\subsubsection{In-Processing Price-Sensitivity Methods}
\label{sec:price_sens_inproc}

Most of the approaches used to generate price-sensitive recommendations are based on in-processing algorithms.
The main characteristic of these algorithms is that price sensitivity is incorporated directly into the model.

In particular, this methodology proved particularly flexible to be applied to the well-known \textit{Matrix Factorization} (\textit{MF}) \cite{koren_advances_2021, koren_matrix_2009} model.
The original model estimates the expected interest
of the user toward a given item
via the dot product of
latent factor vectors.
These are traditionally learned through a dimensionality reduction algorithm applied to the user-item interaction matrix.
Considering price-sensitive methodologies based on MF, for example, one paper \cite{ge_cost-aware_2014} proposes to incorporate cost factors\footnote{Note that here we respect the original paper's terminology by referring to the cost, but actually the cost for the user is simply the item's price.} into the model's objective function to generate more accurate travel tour recommendations.
The experiments reported by the authors indicate that explicitly incorporating cost factors
improves the overall accuracy of the recommendations when compared with a plain MF model.
Also, extending MF, other papers in the literature \cite{chen_does_2014, chen_boosting_2017} propose incorporating customers' price preferences explicitly into the objective function through the use of particular regularizers.
However, whereas previously the purpose was to enhance the overall performance of the system, here the study is about the use of price preferences to make recommendations in product categories that the user has never explored (\textit{transfer learning}).
In particular, according to the authors, generating recommendations for customers' unexplored product categories can cause significant performance drops (-$40\%$) if traditional algorithms are used, since the learned product user preferences are difficult to transfer from one category to another.
Instead, explicitly incorporating customers' price preferences into the objective function can help to significantly improve (+$43\%$) performance on unexplored categories compared to state-of-the-art baselines.

Other studies in the literature \cite{greenstein-messica_personal-discount_2017, greenstein-messica_personal_2018, umberto2015developing}
propose incorporating customers' price preferences within existing \emph{context-aware} recommendation algorithms \cite{kulkarni_context_2020}.
In particular, an experimental study \cite{umberto2015developing} proposed adapting collaborative filtering algorithms by integrating price as an additional dimension of the well-known multi-dimensional context-aware recommendation model \cite{adomavicius2010context}.
As the study found, the integration of price information allows the accuracy of recommendations to be increased on average.
However, the way the price is set could impact business performance.
Indeed, a price-sensitive RS could increase the accuracy of predicting cheap items
and simultaneously decrease the accuracy of predicting more expensive items, thus yielding negative effects on corporate profitability.
Similar results were also found by another study with real customers in the food \& beverage field \cite{greenstein-messica_personal-discount_2017}.
According to the study, explicitly incorporating discount sensitivity into the algorithms can help to significantly improve performance in a coupon recommendation task when compared to the CAMF method \cite{baltrunas_matrix_2011}, i.e., a context-aware variant of matrix factorization.
Specifically, in the domain of location-based deals, the analysis shows that the most important feature for predicting purchase probability is the discount-to-distance ratio:
the higher the discount offered by the store, the more likely the customer is to travel longer distances to obtain it.
However, as is well known in the literature, context variables often depend on the considered business domain.
In particular, eBay.com has some unique characteristics \cite{greenstein-messica_personal_2018}.
In this multi-seller platform, the same products are offered at various prices simultaneously by various sellers with different reputation scores.
According to a study \cite{greenstein-messica_personal_2018}, in this business domain, incorporating customers' \textit{willingness-to-pay} (\textit{WTP}), discounting, and seller reputation features into a context-aware recommender can help to significantly improve the accuracy of predictions, with an $84\%$ improvement over MF models.

In addition, recent studies \cite{zheng_price-aware_2020, zheng_incorporating_2021, zhang_price_2022} propose incorporating customers' price preferences into algorithms by exploiting Graph Neural Networks (GNNs) \cite{gao_survey_2023}.
Specifically, in two related studies \cite{zheng_price-aware_2020, zheng_incorporating_2021}, it is proposed to construct a GNN-based recommender by building a heterogeneous graph consisting of different types of nodes: customers, items, prices, and product categories.
The key idea is to propagate price influence from prices to users by leveraging items as a bridge so that price preferences are implicitly encoded into the embeddings.
The use of price-sensitive GNNs is also exploited in the field of session-based recommendations \cite{zhang_price_2022}.
For all studies based on GNNs \cite{zheng_price-aware_2020, zheng_incorporating_2021, zhang_price_2022}, the models are able to generate slightly more relevant recommendations than the baselines.
However, as various authors pointed out, it is difficult to handle heterogeneous information and model complex relationships underlying customer buying behavior, and research still offers many opportunities to develop better-performing models that can fully exploit the potential of GNN-based algorithms.

\subsubsection{Post-Processing Price-Sensitivity Methods}

A number of price-sensitive recommendation algorithms also make use of post-processing methods.
The latter are primarily re-ranking algorithms, which can be applied on top of any price-agnostic recommender baseline.

\begin{table}[tp]

\footnotesize
	\begin{center}
    \begin{tabular}{m{0.5cm} m{7.5cm} }
    \toprule
    Ref & Re-Ranking Method \\
    \midrule
    \cite{zhang_identifying_2018} &
    \vbox{\begin{equation}
    \underset{\mathcal{Y}_{u,k}}{\mathrm{argmax}}
    \sum_{i \in \mathcal{Y}_{u,k}}{ \hat{x}_{u,i}(\mathbf{\Theta}) \cdot s_{u,i} (\mathbf{\Phi}) }
    \label{eq:price_sens_1}
    \end{equation} \vspace{-0.5cm}} \\

    \cite{wadhwa_personalizing_2020}* &
    \vbox{\begin{equation}
    \underset{\mathcal{Y}_{u,k}}{\mathrm{argmax}}
    \sum_{i \in \mathcal{Y}_{u,k}}
    w_1(\mathbf{\Psi}) \cdot \hat{x}_{u,i}(\mathbf{\Theta}) + w_2(\mathbf{\Psi}) \cdot s_{u} (\mathbf{\Phi})
    \label{eq:price_sens_2}
    \end{equation} \vspace{-0.8cm}} \\

    \cite{kompan_exploring_2022} &
    \vbox{\begin{multline}
    \underset{\mathcal{Y}_{u,k}}{\mathrm{argmax}}
    \sum_{i \in \mathcal{Y}_{u,k}} \hat{x}_{u,i}(\mathbf{\Theta}) \cdot \Biggl( \Bigl(1 + \log_{10}\bigl(0.1 + \frac{0.9 \cdot p_i}{c_i}\bigl) \Bigl)^\beta + \\
    + \Bigl(1 + \log_{10}\bigl(0.1 + \frac{0.9 \cdot p_i}{\bar{\mathbf{p}}^u}\bigl) \Bigl)^\gamma \Biggl)
    \label{eq:price_sens_4}
    \end{multline} \vspace{-0.5cm}} \\

    \bottomrule
    \end{tabular}
    \end{center}
    \vspace{-0.3cm}
    \caption{Price-sensitive re-ranking methods.
    *The formula captures the main essence of the described approaches.}
    \vspace{-0.5cm}
	\label{tab:price_sensitive_rankers}
\end{table}

In this domain, it is proposed, for example, to generate recommendations by weighting the expected interest $\hat{x}_{u,i}(\mathbf{\Theta})$ by the price-sensitivity $s_{u,i}(\mathbf{\Phi})$.
The latter is a particular variable, learned through a different model parameterized by $\mathbf{\Phi}$, indicating how price-sensitive a given user is to a given item (Eq.~\ref{eq:price_sens_1}) \cite{zhang_identifying_2018}.
A similar approach is also proposed in another study \cite{wadhwa_personalizing_2020}.
However, in this case, the price-sensitivity variable $s_{u}(\mathbf{\Phi})$ depends only on the customer and not on the item (Eq.~\ref{eq:price_sens_2}).
In addition, it is necessary to use another regression model (parameterized by $\mathbf{\Psi}$) to learn how to properly weigh (through $w_1(\mathbf{\Psi})$, $w_2(\mathbf{\Psi})$ coefficients) the price-sensitivity with the user's expected interest.
Both studies show that through the use of price-sensitivity algorithms, more relevant recommendations can be obtained.

Recently, another research \cite{kompan_exploring_2022} proposes a hybrid approach (Eq.~\ref{eq:price_sens_4}) combining the price-sensitive and the profit-aware\footnote{We discuss profit-aware methods in Section~\ref{sec:profit_aware}.} subdomains.
This approach weighs the expected interest $\hat{x}_{u,i}(\mathbf{\Theta})$ by balancing a user price preference factor $\frac{p_i}{\bar{\mathbf{p}}^u}$ with a profitability factor $\frac{p_i}{c_i}$, where $\beta, \gamma \in [-1, 1]$ in Eq.~\ref{eq:price_sens_4} are regularization parameters. In particular, considering $\bar{\mathbf{p}}^u$ as the average user price, the first factor captures the difference between the customer's typical price level and the actual item's price.
The second factor, $\frac{p_i}{c_i} = 1 + \frac{v_i}{c_i}$, captures how much an item's sale is able to repay the underlying cost and bring profit to the organization.
In this way, it becomes possible to effectively balance the interests of customers with those of the organization because the increase in profitability that traditionally adversely affects the relevance of recommendations is more than offset by the increase in the latter due to the influence of price preferences.

In Table~\ref{tab:price_sensitive_rankers}, we formally characterize the three discussed price-sensitive re-ranking methods.

\subsection{Economic Utility Modeling Methods}
\label{sec:utilitarian}

In the economic literature \cite{mcconnell2020macroeconomics}, user behavior is often modeled using utilitarian theories to construct systems that can describe and/or optimize certain dynamics.
According to the \textit{Rational Choice Theory} (\textit{RCT}), at each time instant, a rational user, when faced with a set of alternatives, will choose those with the highest utility for him or her \cite{scott2000rational}.
Accordingly, many studies in the literature \cite{wang_utilizing_2011, ge_maximizing_2019, zhao_multi-product_2017}  propose algorithms that explicitly consider the customer's utilitarian behavior to generate more useful recommendations that can in turn increase conversion rates and profitability.
Below we give some insights on how these methods work by discussing a few selected articles focused, respectively, on multi-attribute, repurchase, and complementary recommendations.

\begin{table}[t]
\footnotesize
	\begin{center}
    \begin{tabular}{m{0.5cm} m{2.5cm} m{7cm}}
    \toprule
    Ref & Name & Utility Function \\
    \midrule
    \cite{wang_utilizing_2011} & Standard &  \vbox{
    \begin{equation}
    \begin{aligned}
    & \mathcal{T}(\mathcal{Y}_{u,k}) = \sum_{i \in \mathcal{Y}_{u,k}} \rho_{u,i}
    \end{aligned}
    \label{eq:utility_1}
    \end{equation}
    \vspace{-0.8cm}} \\

    \cite{huang_designing_2011} & Multi-Attribute &  \vbox{
    \begin{equation}
    \begin{aligned}
    & \mathcal{T}(\mathcal{Y}_{u,k}) = \sum_{i \in \mathcal{Y}_{u,k}} \sum_{g \in \mathcal{G}} f_{u,g} \cdot \rho_{i,g}
    \end{aligned}
    \label{eq:utility_10}
    \end{equation}
    \vspace{-0.8cm}} \\

    \cite{wang_utilizing_2011}* & Constant Elasticity of Substitution &  \vbox{
    \begin{equation}
    \begin{aligned}
    & \mathcal{T}(\mathcal{Y}_{u,k}) = \sum_{i \in \mathcal{Y}_{u,k}} \rho_{u,i} \cdot q_{u,i}^{\xi_i}
    \end{aligned}
    \label{eq:utility_4}
    \end{equation}
    \vspace{-0.8cm}} \\

    \cite{zhang_economic_2016} & King-Plosser-Rebelo &  \vbox{
    \begin{equation}
    \begin{aligned}
    & \mathcal{T}(\mathcal{Y}_{u,k}) = \sum_{i \in \mathcal{Y}_{u,k}} \rho_{u,i} \cdot \ln(1 + q_{u,i})
    \end{aligned}
    \label{eq:utility_7}
    \end{equation}
    \vspace{-0.8cm}} \\

     \cite{zhao_multi-product_2017} & Multi-Product &  \vbox{
    \begin{equation}
    \begin{aligned}
    & \mathcal{T}(\mathcal{Y}_{u,k}) = \frac{1}{|\mathcal{Y}_{u,k}|} \sum_{i,j \in \mathcal{Y}_{u,k}: i \neq j} \Bigl(  a_{i,j} \cdot q_{u,i}^{1 - b_{i,j}} + \\
    & \qquad + (1 - a_{i,j})\cdot q_{u,j}^{1 - b_{i,j}} \Bigr)^{\frac{1}{1- b_{i,j}}}
    \end{aligned}
    \label{eq:utility_8}
    \end{equation}
    \vspace{-0.8cm}} \\

    \cite{ge_maximizing_2019} & Marginal Utility per Dollar &  \vbox{
    \begin{equation}
    \begin{aligned}
    & \mathcal{T}(\mathcal{Y}_{u,k}) = \sum_{i \in \mathcal{Y}_{u,k}} \frac{\tanh{(\rho_{u,i})} \cdot r_{i,u}}{(1 + q_{u,i})\cdot \sigma(p_i)}
    \end{aligned}
    \label{eq:utility_9}
    \end{equation}
    \vspace{-0.5cm}} \\

    \bottomrule
    \end{tabular}
    \end{center}
    \vspace{-0.3cm}
    \caption{Economic utility functions from rational choice theory.
    *The formulas capture the main essence of the described approaches.
    \vspace{-0.5cm}}
	\label{tab:utilitarian_formula}
\end{table}

In the field of RSs, many studies in the literature assume that the utility $\rho_{u,i}$ of a product to a customer depends on his or her purchase history \cite{wang_utilizing_2011}.
Most existing RSs recommend for each user $u$ a list $\mathcal{Y}_{u,k}$ consisting of the top-$k$ items (Eq.~\ref{eq:rerank_base}) with the highest predicted scores $\hat{x}_{u,i}(\mathbf{\Theta})$.
The list $\mathcal{Y}_{u,k}$ is traditionally selected from a set of items with which the user has never interacted before.
Interpreting this assumption from the perspective of economic utility theory (Eq.~\ref{eq:utility_1}) \cite{wang_utilizing_2011}, then, the utility $\mathcal{T}(\mathcal{Y}_{u,k})$ of a recommendation $\mathcal{Y}_{u,k}$
is nothing but the sum of the predicted scores, i.e., $\rho_{u,i} = \hat{x}_{u,i}(\mathbf{\Theta})$.
In this case, a recommendation $\mathcal{Y}_{u,k}$ generated by optimizing the total utility of a set of $k$ recommended items optimizes the expected user interest estimated by any recommendation algorithm.

However, in addition to the previous utility definition,
alternative definitions are recently emerging in the literature.
For example, in the field of \textit{Multi-Criteria Recommendation Systems} (\textit{MCRS}) \cite{al-ghuribi_multi-criteria_2019}, in the presence of a set $\mathcal{G}$ of attributes associated with items, various studies in the literature \cite{huang_designing_2011, dorner_predicting_2013, deng_utility-based_2015, scholz_measuring_2015, zheng_utility-based_2019} propose to generate recommendations by exploiting the \textit{Multi-Attribute Utility Theory} (\textit{MAUT}) \cite{keeney1993decisions}.
MAUT is one of the most widely used utilitarian theories in decision making, which aims to weigh a set of relevant variables to determine the overall utility of each alternative.
In the context of recommendations, in particular, the overall optimized utility (Eq.~\ref{eq:utility_10}) in this case depends on the utility $\rho_{i,g}$ of the single attribute $g$ of item $i$, and a weight $f_{u,g}$ that each user can provide to indicate the importance of that attribute.

Other studies focus on the problem of repeated purchase recommendations \cite{wang_utilizing_2011, zhang_economic_2016, zhao_multi-product_2017, ge_maximizing_2019}.
Unlike traditional RSs, algorithms developed for this task generate recommendations by also considering items that the user already purchased in the past.
In particular, it is observed that the repurchase cycle of some products may follow the \textit{Law of Diminishing Marginal Utility} \cite{wang_utilizing_2011, mcconnell2020macroeconomics}.
According to this theory, many products have decreasing utility for the user as the quantity of purchased products increases (e.g., computers, cell phones), while others, instead, are likely to be purchased frequently over time (e.g., baby diapers, pet food).
Using the standard utilitarian criterion in Eq.~\ref{eq:utility_1} it is not possible to model this behavior.
Indeed, in this case, the usefulness of recommendations for the user highly depends on the quantity $q_{u,i}$ of item $i$ purchased by him or her until a specific time.
In this context, promising results can be obtained by modeling the repurchase cycle through the \textit{Constant Elasticity of Substitution Utility Function} \cite{uzawa1962production}.
This allows the decreasing marginal utility of product $i$ to be properly modeled through a parameter $\xi_i \in [0, 1]$ associated with item $i$ (Eq.~\ref{eq:utility_4}).
This parameter can be learned by extending the MF objective function.
In this way, the algorithm can explicitly consider the decreasing utility of certain products for the user and generate more relevant recommendations.

With similar methodologies, other utilitarian functions are also used in the literature to model customer purchasing behavior \cite{zhang_economic_2016, zhao_multi-product_2017, ge_maximizing_2019}.
However, these studies focus on different objectives.
For example, one study \cite{zhang_economic_2016} proposes three different business cases (i.e., e-commerce, P2P lending, freelancing) that exploit the
\textit{King-Plosser-Rebelo Utility Function} (Eq.~\ref{eq:utility_7}) to optimize the \textit{Total Surplus}, i.e., an indicator that considers both the usefulness of the recommendations for the customer and the profit for the producer.
Another study \cite{zhao_multi-product_2017} in contrast propose to use the \textit{Multi-Product Utility Function} (Eq.~\ref{eq:utility_8}) in order to also consider any complementary and substitutability relationships among the recommended products\footnote{Note that alternative approaches for generating complementary recommendations that are not based on particular utility theories and that do not explicitly optimize business KPIs are available in the literature, e.g., \cite{hao2020p, zhang2018quality, sun2017exploiting}.}.
In the equation, the variables $a_{i,j}$ and $b_{i,j}$ are additional parameters that the recommendation algorithm can jointly learn with the latent factors
to model the indifference curves between pairs of products, i.e., how much the increase in one product affects the relative marginal utility of another product.
Finally, one study \cite{ge_maximizing_2019} proposes using the \textit{Marginal Utility per Dollar Function}
(Eq.~\ref{eq:utility_9}).
This function considers the price $p_i$ of item $i$ and a risk attitude coefficient $r_{i,u}$ to model customers' risk-aversion, i.e., the tendency of consumers to spend only a small portion of their total wealth on a single purchase.

In Table~\ref{tab:utilitarian_formula}, we formally characterize the utility criteria discussed above.

\subsection{Profit-Aware Methods}
\label{sec:profit_aware}

Profit is one of the most important business KPIs for a successful enterprise \cite{teece_internal_1981}.
Accordingly, many studies in the literature \cite{chen_developing_2008, kompan_exploring_2022, goyal_recmax_2012, li_revman_2021, nemati_devising_2020, concha-carrasco_multi-objective_2023} propose profit-aware recommendation algorithms to directly optimize the firm's profitability.
Below we give some insights on how these methods work by discussing a few selected articles.

\subsubsection{In-Processing Profit-Aware Methods}

Profit-aware in-processing approaches in the literature are quite heterogeneous, scattered, and there are several parallel lines of research.
Below, we offer a brief overview of major research directions in this area.

Some early studies \cite{piton_capre_2011, huang_incorporating_2013, yang_real_2017} exploit \textit{Association Rules} \cite{hipp_algorithms_2000}.
According to this particular methodology \cite{park_literature_2012}, recommendations are generated through a frequentist approach based on statistical support and confidence constructs \cite{osadchiy_recommender_2019}.
One of Amazon's most prominent recommenders, i.e., \quotes{\textit{customers who bought this item also bought}}, is seemingly based on association rules.
In particular, many studies in the literature \cite{goos_profit_2002, wang_item_2002, wang_mining_2005, kobti_profit-based_2007} propose to generate association rules while also optimizing profitability.
The main methods incorporate profit considerations when weighting the rules \cite{cai_mining_1998}.
However, unlike modern RSs based on collaborative and content-based filtering algorithms, association rules \cite{goos_profit_2002} are not personalized, i.e., different users do not get different recommendations.
In addition, association rules may generally face challenges when the total number of recommendable items is very large.

Other studies \cite{brand_random_2005, akoglu_valuepick_2010, qu_cost-effective_2014, li2017utility} propose graph-based approaches.
In particular, two works \cite{akoglu_valuepick_2010, li2017utility} focus on social networks.
The proposed algorithms are designed to explicitly optimize the value of link recommendations.
For example, one of those works \cite{li2017utility}, which considered
not only the likelihood of a potential link for the user but also the value of that link for a social network operator in economic terms (i.e., revenues and costs), demonstrated promising results in optimizing the platform's revenue streams from advertising.
Another relatively recent approach based on graphs \cite{qu_cost-effective_2014} is developed specifically for the taxi industry.
In this particular application domain, if we assume an hourly rate, a taxi driver's profit depends solely on the hours billed to customers:
simply put, it is critical for a taxi driver to minimize the distance to find a customer and maximize the distance traveled with a customer on board.
The proposed algorithm recommends pick-up points for taxi drivers in order to maximize the profit of driving routes while balancing the potential congestion resulting from multiple requests from different customers at the same location.

More recently, a study \cite{cai_trustworthy_2019} proposes a profit-aware RS based on collaborative filtering.
The algorithm is based on an extension of the well-known neighbor selection criterion of the user-based nearest-neighbor collaborative filtering model \cite{ricci_trust_2022}.
The original algorithm calculates the predicted score
based on a weighted sum of similarities between users belonging to a given neighborhood.
usually based on correlation criteria.
The authors of \cite{cai_trustworthy_2019} instead propose to calculate the predicted scores
by selecting the neighbors
that would allow the generation of the highest value-weighted expected interest.
Although the focus of the paper is on shilling attacks, i.e., attacks by malicious users who generate biased ratings to influence recommendations for their interests, the subprocedure for selecting the most valuable users can be used to generate more profitable recommendations.

Other recent approaches \cite{wu_turning_2018, lin_pareto-efficient_2019, li_revman_2021} are based on \textit{Learning To Rank} (\textit{LTR}) \cite{chuan_he_survey_2008}.
This is a well-known technique in \textit{Information Retrieval} (\textit{IR}) \cite{kobayashi_information_2000, said_information_2019}.
IR algorithms aim to help users to find the most relevant items based on specific search queries.
In particular, one study \cite{wu_turning_2018} uses this methodology in a product search application context.
The proposed algorithm integrates the price into the objective function in order to optimize the overall sales revenue of an e-commerce.
This technique is later applied \cite{lin_pareto-efficient_2019, li_revman_2021} also to generate recommendations without the need to anchor them to an underlying search query.
For example, one study \cite{lin_pareto-efficient_2019} proposes a multi-objective algorithm that is able to optimize multiple objective functions simultaneously through LTR.
The algorithm is Pareto-efficient, i.e., it optimizes each objective (e.g., CTR and GMV\footnote{We provide the definition of the most frequently used online metrics in Table~\ref{tab:online_metrics}.}) one at a time, with the constraint that no single objective can be further improved without affecting others.

Finally, some studies \cite{nemati_devising_2020, concha-carrasco_multi-objective_2023} propose using profit-aware multi-objective evolutionary algorithms \cite{horvath_evolutionary_2017, zheng_survey_2022}.
One of these \cite{nemati_devising_2020} is based on \textit{Non-dominated Sorting Genetic Algorithm II} (\textit{NSGA-II}).
A more recent one \cite{concha-carrasco_multi-objective_2023} is based on \textit{Multi-Objective Artificial Bee Colony} (\textit{MOABC}).
In these cases, the optimization target is a combination of the item's profit and the user's expected interest.
Both algorithms obtained very promising offline results on the overall profit improvement, although the comparison was performed exclusively with a traditional user-based collaborative filtering algorithm \cite{ricci_trust_2022}.

\subsubsection{Post-Processing Profit-Aware Methods}

\begin{table}[tp]
\footnotesize
	\begin{center}
    \begin{tabular}{m{0.5cm} m{9cm} }
    \toprule
    Ref & Re-Ranking Method \\
    \midrule

    \cite{chen_developing_2008} &
    \vbox{\begin{equation}
    \underset{\mathcal{Y}_{u,k}}{\mathrm{argmax}}
    \sum_{i \in \mathcal{Y}_{u,k}}{\hat{x}_{u,i} (\mathbf{\Theta}) \cdot v_i}
    \label{eq:profit_rerank_1}
    \end{equation}\vspace{-0.9cm}} \\

    \cite{das_maximizing_2009} &
    \vbox{
    \begin{equation}
    \begin{aligned}
    & \underset{\mathcal{Y}_{u,k}}{\mathrm{argmax}}
    \sum_{i \in \mathcal{Y}_{u,k}}{\hat{x}_{u,i} (\mathbf{\Theta}) \cdot v_i} \\
    & \text{s.t. } \text{Dice }(\hat{\mathbf{x}}_{u} (\mathbf{\Theta}), \hat{\mathbf{x}}_{u} (\mathbf{\Theta})^\intercal \mathbf{v}) \ge \eta
    \end{aligned}
    \label{eq:profit_rerank_2}
    \end{equation}
    \vspace{-0.9cm}} \\

    \cite{wang_strategy-oriented_2012}* &
    \vbox{
    \begin{equation}
    \begin{aligned}
    & \underset{\mathcal{Y}_{u,k}}{\mathrm{argmax}}
    \sum_{i \in \mathcal{Y}_{u,k}}{v_i} \\
    & \text{s.t. } \hat{x}_{u,i} (\mathbf{\Theta}) \ge \zeta, \quad p_{i} \le \lambda_u
    \end{aligned}
    \label{eq:profit_rerank_3}
    \end{equation}
    \vspace{-0.9cm}}\\

    \cite{jannach_price_2017}* &
    \vbox{
    \begin{equation}
    \begin{aligned}
    & \underset{\mathcal{Y}_{u,k}}{\mathrm{argmax}}
    \sum_{i \in \mathcal{Y}_{u,k}}{\hat{x}_{u,i} (\mathbf{\Theta}) \cdot v_i} \\
    & \text{s.t. } \hat{x}_{u,i} (\mathbf{\Theta}) \ge \zeta
    \end{aligned}
    \label{eq:profit_rerank_4}
    \end{equation}
    \vspace{-1.1cm}} \\

    \cite{ghanem_balancing_2022} &
    \vbox{\begin{equation}
    \underset{\mathcal{Y}_{u,k}}{\mathrm{argmax}}
    \sum_{i \in \mathcal{Y}_{u,k}}{\delta \cdot \hat{x}_{u,i}(\mathbf{\Theta}) + (1 - \delta) \cdot v_i}
    \label{eq:profit_rerank_5}
    \end{equation}\vspace{-1.1cm}} \\

    \cite{malthouse_multistakeholder_2019}* &
    \vbox{
    \begin{equation}
    \begin{aligned}
    & \underset{\mathcal{Y}}{\mathrm{argmax}}
    \sum_{u \in \mathcal{U}}\sum_{i \in \mathcal{I}}{\delta \cdot y_{u,i} \cdot \hat{x}_{u,i}(\mathbf{\Theta}) + (1 - \delta) \cdot y_{u,i} \cdot d_{u,i}} \\
    & \text{s.t. } \sum_{i \in \mathcal{I}} \mathbbm{1}(y_{u,i} \cdot d_{u,i} > 0) \le l \quad (\forall u \in \mathcal{U}) \\
    \end{aligned}
    \label{eq:profit_rerank_6}
    \end{equation}
    \vspace{-0.9cm}} \\

    \bottomrule
    \end{tabular}
    \end{center}
    \vspace{-0.3cm}
    \caption{Profit-aware re-ranking methods.
    *The formulas capture the main essence of the described approaches.
    \vspace{-0.5cm}}
	\label{tab:profit_awareness_rankers}
\end{table}

In the context of this survey, many profit-aware approaches rely on post-processing re-ranking methods.
As mentioned earlier, these approaches consider the recommender baseline as a black box and generate recommendations by exploiting a combination of certain heuristics.

All examined profit-aware approaches are based on a simple but important assumption \cite{jannach_price_2017, debiasio_value_2023}:
the items most relevant to the user are often not those of the highest business value to the organization.
Consequently, prioritizing the highest-profit items in recommendations would allow for increased business profitability as a result of actual user purchases of those items.
In one of the earliest approaches \cite{mu-chen_chen_hprs_2007, chen_developing_2008} it is proposed to weight the probability of purchase (i.e., the estimated expected interest) by profitability in order to maximize an average expected profit (Eq.~\ref{eq:profit_rerank_1}).
This approach should make it possible to provide more profitable recommendations than those generated by traditional RSs.
Experiments in a synthetic dataset based on a subset of groceries transactions show encouraging results:
the proposed algorithm was able to increase profitability without excessively impacting the relevance of recommendations.
However, as also reported by the authors \cite{mu-chen_chen_hprs_2007, chen_developing_2008}, the interests of customers and the organization must be balanced appropriately.
In fact, the organization could risk losing loyal customers should they feel dissatisfied with overly biased recommendations toward higher-value items and decide to leave the platform.

To mitigate this drawback, and thus to avoid providing completely irrelevant recommendations, various studies propose more or less straightforward extensions of Eq.~\ref{eq:profit_rerank_1} based on constrained optimization techniques.
One of the earliest papers \cite{das_maximizing_2009} proposes a constrained re-ranking method based on the \textit{Dice} coefficient (Eq.~\ref{eq:profit_rerank_2}).
This can help prevent the system from providing recommendations that are too dissimilar from the original ones based on a threshold $\eta$.
However, the study is based on various simplifying assumptions and does not provide an empirical evaluation of the approach.
In two related studies \cite{wang_mathematical_2009, wang_strategy-oriented_2012} instead, it is proposed to maximize profitability under customer satisfaction and budget constraints (Eq.~\ref{eq:profit_rerank_3}), where $\zeta$ and $\lambda_u$ are two thresholds used to keep the probability of purchase and the price of items within certain ranges, respectively.
In particular, an expert system is proposed where different optimization goals can be specified in order to optimize profitability or balance profitability and satisfaction in order to achieve a win-win situation for suppliers and customers.
A similar variant of this approach (Eq.~\ref{eq:profit_rerank_4},~\ref{eq:profit_rerank_5}) is also proposed in two related studies \cite{jannach_price_2017, ghanem_balancing_2022} where the short- and long-term profit-relevance tradeoff is investigated through the use of simulations.
In Eq.~\ref{eq:profit_rerank_5}, $\delta \in [0,1]$ is a regularization parameter.

In addition, other studies \cite{malthouse_multistakeholder_2019, malthouse_algorithm_2019, zhang_smart_2017} propose algorithms to address the problem of sponsored recommendations.
In this scenario, a supplier who decides to sponsor its products pays the platform for each user interaction.
One study \cite{malthouse_multistakeholder_2019} in particular proposes a multi-objective post-processing re-ranking algorithm (Eq.~\ref{eq:profit_rerank_6}).
In the equation, $y_{u,i}$ is a decision variable ($y_{u,i} = 1$ iff $i \in \mathcal{Y}_{u,k}$), $d_{u,i}$ is the ad revenue that the organization gets from suppliers if the user interacts with the item, and $l < k$ is the maximum number of sponsored items that can be included in the recommendation list.
The algorithm is designed to balance the recommendation of high ad revenue sponsored items with the user's interests.

In Table~\ref{tab:profit_awareness_rankers}, we formally characterize the profit-aware re-ranking methods discussed above.

\subsection{Promotional Methods}
\label{sec:promotion}

Promotional
methods \cite{morgan_research_2019, anagnostopoulos_recommendation_2013} aim to increase sales figures by
promoting products and services to the most appropriate customer segments.
We identify three main strategies in the RSs literature that can be used to optimize profit and related business KPIs.
\textit{Pricing methods}, can be used to offer products at a discounted price or to strategically adjust prices in order to increase the market demand.
\textit{Bundling methods}, are special pricing methods that are applied to product bundles.
\textit{Brand-awareness methods}, finally, can be used to focus customers' attention on the organization's products in order to generate extra sales.
Below we give some insights on how these methods work by discussing a few selected articles in each category.

\subsubsection{Pricing Methods}
\label{sec:dyn_pricing}

As discussed earlier in Section~\ref{sec:price_sens}, price is one of the most influential variables of customer buying behavior, and considering this variable explicitly would allow for recommendations more in line with customers' interests.
However, while the previous section focuses on customer-oriented methodologies that integrate price sensitivity as additional information in order to generate more relevant recommendations, in this section we instead discuss promotional techniques that an organization might want to apply to incentivize the purchase of certain products by strategically setting the prices \cite{bergemann_optimal_2006, ghoshal_recommendations_2021}.
In the following, we describe two organizational strategies referring to:
(a) \textit{occasional discounting}; (b) \textit{personalized dynamic pricing}.

One of the most commonly used promotional strategies to incentivize product purchases is to offer occasional discounts \cite{chen_effects_1998}, for example at certain times of the year (e.g., winter sales) or special events (e.g., Black Friday).
In the context of RSs many studies \cite{wu_enhanced_2011, jannach_determining_2017, wang_personalized_2020, sato_discount_2015, jiang_optimization_2012, jiang_redesigning_2015, guo_maximizing_2020} aim to generate recommendations while considering discounts.
Some studies propose, for example, to use re-ranking algorithms \cite{jannach_determining_2017} to promote products on sale or in-processing methods \cite{wang_personalized_2020} based on adaptations of MF-based models
to explicitly consider customers' discount sensitivity \cite{sato_discount_2015}.
Another method is proposed in two related studies \cite{jiang_optimization_2012, jiang_redesigning_2015}.
In particular, as noted by the authors, there may be inter- and cross-category effects when discount products are bought.
Thus, especially in e-commerce, organizations can exploit RSs to incentivize customers to buy discount products but also those products that are related to them but not on sale
(e.g., camera on sale and full-price lens).
A similar analysis is also made \cite{guo_maximizing_2020} to determine the optimal shipping-fee discount to attract customers to the platform and encourage them to purchase products related to the discounted ones.

While discounts may be occasional and the same for all customers, some methodologies are proposed in the RSs literature to generate dynamic customer-specific prices in order to strategically promote certain products and generate higher profits.
In this context, some initial studies \cite{backhaus_incorporating_2010, backhaus_enabling_2010} propose to use survey-based techniques (\textit{conjoint analysis})
to estimate customer WTP (recall Section~\ref{sec:price_sens_inproc}) and filter items that are priced higher than WTP in the ranking.
The authors also discuss some possible configurations of the algorithm to set the prices based on WTP in order to generate more profit for the organization.
However, the proposed pricing model is only theoretical as it is not validated by empirical experiments.
Another study \cite{kamishima_personalized_2011} proposes a system that classifies customers based on whether they would buy products only if discounted or not.
Based on the type of customer, the system can offer a discount in order to incentivize purchases.
However, as discussed later \cite{massoud_conceptual_2012}, the study is based on assumptions that are not feasible in practice:
all products have the same price;
only two price values are available (i.e., standard and discounted price).
Another work \cite{zhao_e-commerce_2015} proposes a different methodology.
The study focuses on a lottery-based mechanism that aims to obtain the exact WTP for one subset of products and then to exploit this information to predict the WTP for another subset of products.
In this way, the system can offer a personalized promotion to increase the conversion rate of the latter product subset.
The authors report significant results on the potential ability of this system to increase profit over conventional systems.
However, the experiments are based on a user study with a low number of users.
Finally, another study \cite{adelnia_najafabadi_dynamic_2022} proposes a dynamic personalized pricing RS for information goods (e.g., digital movie rentals).
These goods differ from physical goods in that their production and distribution costs are negligible and they can be copied, rented, and resold easily.
In this context, traditional markup-based pricing methods (i.e., cost plus margin) are not effective because there is no true underlying unit cost.
The proposed system first classifies customers according to their WTP and quality sensitivity (e.g., whether they prefer a premium version of the same product).
Then it calculates a personalized price to incentivize purchase.

\subsubsection{Bundling Methods}
\label{sec:bundling}

One frequently used promotional strategy \cite{rao_design_2009} to increase sales revenue of certain products
is to offer them at a discount if purchased in bundles \cite{harlam_impact_1995}.
In the literature \cite{yan_profit_2011} it is proposed, for example, to include in the bundles:
(a) products that are complementary to each other in order to incentivize cross-selling;
(b) products that are uncorrelated, for example, to clear the stock in the warehouse;
(c) the same product in multiple quantities (e.g., 2x1 promotion).
Specifically, in RSs research \cite{chen_survey_2020}, one branch of the literature focuses on
recommending bundles to optimize profit
by exploiting price modeling techniques.
The other branch, in contrast, does not exploit such techniques and focuses solely on optimizing relevance\footnote{Relevance-based bundling algorithms \cite{sun_revisiting_2022} can be based for example on association rules \cite{yang_comparison_2006, fang_customized_2018, jiao_product_2005}, graph-based approaches \cite{liu_modeling_2017, ge_effects_2017, liu_modeling_2017, deng_personalized_2020, bai_personalized_2019}, GNNs \cite{chang_bundle_2020, agarwal_improving_2022, chang_bundle_2023} and transformers \cite{avny_brosh_bruce_2022}.}.
In this review, we focus only on bundling approaches that aim to explicitly optimize business KPIs.

Concerning price modeling bundle recommendation techniques, two related earlier studies \cite{garfinkel_design_2006, garfinkel_shopbot_2008} focus on the development of a shopbot (i.e., comparison shopping agents) capable of offering bundles at a discounted price based on an integer linear programming model.
The proposed algorithm is validated using data from Amazon.com and Buy.com reporting significant results from the perspective of potential economic savings of price-sensitive bundle purchasing customers.
However, the data sample used is very small, and optimization of business KPIs is not explicitly considered.
In contrast, two other studies \cite{jiang_optimizing_2011, zhu_bundle_2014} leverage similar integer programming-based approaches to recommend bundles with the goal of optimizing profitability \cite{jiang_optimizing_2011} or any business objective \cite{zhu_bundle_2014}.
In particular, considering the case where the bundle can be created directly by the customer by selecting the products of his or her preference, the first study \cite{jiang_optimizing_2011} proposes a multistage approach that can dynamically determine the price of the added products in real-time with the goal of maximizing profits for the organization.
In contrast, the second study \cite{zhu_bundle_2014} investigates how to incorporate  product compatibility and potential cost savings to generate bundles that, if recommended, could optimize certain business objectives (e.g., profitability, revenue, and others).
Both studies report results regarding the potential ability of the proposed systems of increasing profitability and conversion rates.
In addition, two other approaches \cite{beladev_recommender_2016, ettl_data-driven_2020} are proposed recently.
The first approach \cite{beladev_recommender_2016} is based on a collaborative filtering algorithm that integrates demand estimation and price modeling techniques to make recommendations
with the goal of jointly maximizing purchase probability and sales revenue considering the customer WTP.
The second approach \cite{ettl_data-driven_2020} is based on an algorithm that can recommend bundles with customized discounts to customers considering also inventory levels.
However, in the former case, the bundle does not offer an additional discount over the full price of the individual products.
Instead, the bundle is created exclusively so that the total price of the products inside it is aligned with the customer's WTP to meet his or her price preferences.
In the latter case, on the other hand, the evaluation is based on a simulation focused on the aviation industry with a large number of assumptions.

\subsubsection{Brand-Awareness Methods}

Some methods in the literature can be used to promote the organization's products and services, raise brand awareness, and increase profitability in the long run.
These methods can be interpreted by referring to the sales funnel \cite{kantola_sales_2020}.
The sales funnel is a theoretical model that describes the customer journey in different stages according to the type of customer interaction with the organization \cite{paschen_collaborative_2020}.
Depending on the status of the customer in the sales funnel, it might be advisable to design an RS with different purposes.

If the customer has not yet made the first purchase (which is referred to as the prospect state), it might be promising to maximize the conversion rate by closing the first deal as quickly as possible \cite{karlsson_using_2013}.
At this early stage, recommending the most popular products may not be the best strategy.
Since many popular products are commonly purchased together, customers would discover them on their own without the need of a recommendation.
Instead, it could be more beneficial to present still popular but unrelated products, optimizing coverage.
In this way, it may be possible to attract more customers to the platform and increase the probability they make their first purchase.

Once the customer has made the first purchase, the company can exploit
mechanisms to optimize profits in the long-run \cite{bodapati_recommendation_2008, goyal_recmax_2012}.
One option could be to mainly recommend
items with high consumer ratings
\cite{jannach2010recommender}.
However, similarly to the previous case, this may not be the best choice either, as many customers might search for and buy such items anyway.
Instead, it might be more valuable to stimulate the purchase of products of possible interest that are likely unknown
to the customer \cite{bodapati_recommendation_2008}, e.g., products that do not fall in the top-$k$ but have medium-high ranking positions.
This way, the company may get both the revenue from the purchases of products that the customer would discover on their own without the recommendations, and an additional revenue through the purchases that were triggered by the recommendations.
With similar objectives, it might also be worthwhile for the company to leverage an RS \cite{goyal_recmax_2012} to launch a marketing campaign with the purpose of promoting new products in the market.
Such a system could be designed to select a set of seed consumers for the marketing campaign such that if these seed consumers provide relatively high ratings, the number of other consumers to whom the new product is recommended is maximized.

\subsection{Long-Term Value Sustainability Methods}
\label{sec:long_term}

It is very important for organizations to grow sustainably over time \cite{lumpkin_long-term_2010, rauch_effects_2005}.
Accordingly, a number of studies in the literature \cite{hosanagar_recommended_2008, iwata_recommendation_2008, pei_value-aware_2019} propose recommendation algorithms that consider temporal dynamics to optimize long-term business value.
Many of them rely on the \textit{Customer Lifetime Value} (\textit{CLV}) \cite{blattberg_customer_2009, bhaduri_customer_2016} and other related conceptual models (e.g., \textit{Recency Frequency Monetary} - \textit{RFM}) from the business literature.
CLV represents the expected business value of all future cash flows attributed to a specific customer discounted to the present time.

Similarly to what is found for bundling methods (in Section~\ref{sec:bundling}), some RSs studies propose to exploit CLV to optimize long-term profit \cite{pei_value-aware_2019, iwata_recommendation_2008} while others exploit it solely to optimize relevance\footnote{Typically, RSs that rely on CLV-related models to optimize relevance \cite{liu_integrating_2005, liu_hybrid_2005, shih_hybrid_2005, shih_product_2008, tabaei_using_2012, park_effective_2015, wang_recommender_2009, pandey_customer_2021, blattberg_customer_2009} follow a common workflow. Algorithms first group users into similar customer value segments. Then they generate recommendations through association rules or collaborative filtering leveraging this additional information.} \cite{shih_product_2008, tabaei_using_2012}.
In this review, we focus only on algorithms that aim to optimize long-term business KPIs.
Below we give some insights on how these methods work by discussing a few selected articles.
In particular, we first discuss in- and post-processing methods based on supervised learning and then we delve into recent algorithms based on reinforcement learning\footnote{Much work has been carried out in the field of reinforcement learning based RSs \cite{afsar_reinforcement_2022}.
However, the majority of these papers, e.g., \cite{liu2021top, liu2020top}, are not the subject of our analysis, as they do not focus on optimizing business KPIs.}.

\subsubsection{Post-Processing and Supervised Learning Methods for Long-Term Business Value Optimization}

Some studies \cite{hosanagar_recommended_2008, hosein_recommendations_2019, basu_personalized_2021, panniello_impact_2016} propose post-processing algorithms to maximize the long-term business value of recommendations by exploiting heuristic criteria.
In particular, \citeauthor{hosanagar_recommended_2008} \cite{hosanagar_recommended_2008} proposes an algorithm following this simple but effective intuition:
when a customer trusts an RS, the system should bias the recommendations to increase profitability;
instead, when the customer trust is below a certain threshold, the system should recommend the most relevant products to restore trust at the expense of profitability.
The original study \cite{hosanagar_recommended_2008} proposes only a theoretical assessment of the profit surplus that can be generated using this algorithm.
However, the algorithm's performance is also evaluated in an online study \cite{panniello_impact_2016} and in a recent post-hoc econometric analysis \cite{basu_personalized_2021}.
These recent studies demonstrate both the effectiveness of the proposed methods in generating higher sales revenue than a content-based filtering algorithm \cite{panniello_impact_2016} and how trust is positively correlated with higher sales revenue \cite{basu_personalized_2021}.

Other approaches based on supervised machine learning algorithms are also studied to explicitly optimize the long-term business value of recommendations.
In particular, in two related studies \cite{iwata_recommendation_2006, iwata_recommendation_2008}, a recommendation system is proposed to explicitly maximize CLV.
The algorithm is designed specifically for subscription-based \cite{iwata_recommendation_2006} and transaction-based \cite{iwata_recommendation_2008} revenue models.
In particular, survival analysis techniques are used to identify frequent purchasing patterns among higher CLV users.
Then, recommendations are generated to match those patterns as closely as possible.
The algorithms are evaluated using real data from a mobile cartoon provider
with a subscription-based revenue model \cite{iwata_recommendation_2006} and an online music provider with a transaction-based revenue model \cite{iwata_recommendation_2008}, both from Japan.
However, although results regarding the improvement of the subscription period and the number of items purchased over time are reported, the evaluation is only based on a simulation system of user purchasing behavior.

\subsubsection{Reinforcement Learning Recommendation Methods for Long-Term Business Value Optimization}

Recent studies propose methodologies based on \textit{Reinforcement Learning} (\textit{RL}) for optimizing the long-term business value of recommendations \cite{sutton2018reinforcement}.
RL is a learning approach that aims to learn an optimal policy (i.e., recommendation strategy) based on the sequential interaction between an agent and the environment through trial and error to maximize a reward.
This methodology is used many times in the literature \cite{zou_reinforcement_2019, zhao_maximizing_2020, ji_reinforcement_2021, guo_we_2021, wu_returning_2017, theocharous_personalized_2015, ju_reinforcement_2017, pei_value-aware_2019, he_propn_2022} to optimize CLV.

A few studies propose algorithms to directly optimize profit \cite{ju_reinforcement_2017, pei_value-aware_2019}.
These studies focus on the transaction-based revenue model where each customer purchase brings a certain profit to the organization.
Specifically, in this context, one study \cite{pei_value-aware_2019} considers that a certain profit share can be allocated to each user action (i.e., click, add-to-cart, pay).
Hence, the overall profitability can be maximized by optimizing the sum of the profit allocated to each user action considering the probability that such an action will occur given the recommendations.
Other studies \cite{zou_reinforcement_2019, zhao_maximizing_2020, ji_reinforcement_2021, guo_we_2021, wu_returning_2017, theocharous_personalized_2015}, in contrast, propose algorithms to optimize user engagement, or more generally some strategic interrelated business indicators \cite{he_propn_2022}.
One study \cite{theocharous_personalized_2015} is based on the advertising revenue model.
In this particular context, advertisers are used to pay the platform a certain monetary amount for each click or conversion generated.
Hence, in this case, by optimizing user engagement, profit is directly optimized.
Instead, other works \cite{zou_reinforcement_2019, zhao_maximizing_2020, ji_reinforcement_2021, guo_we_2021, wu_returning_2017}, although they similarly propose to optimize user engagement, are not based on advertising revenue models.
Therefore, in these cases, the relationship with profitability is indirect, as user engagement positively correlates with retention.

\section{Evaluation Methodologies}
\label{sec:experimental-evaluation}

In this section, we review the evaluation methodologies used in the surveyed papers.
First, we give some insights into the different methodologies that are used to evaluate algorithms.
Next, we discuss the metrics used in offline evaluation.
Then we discuss the results that have been obtained in the real world from ECRSs algorithms by analyzing in detail those studies that report online performance.
Finally, we analyze related topics concerning public datasets and the current level of reproducibility.

\subsection{Evaluation Approaches}

\begin{figure}
\footnotesize
\begin{center}

\includegraphics[height=10.5cm,width=11.5cm]{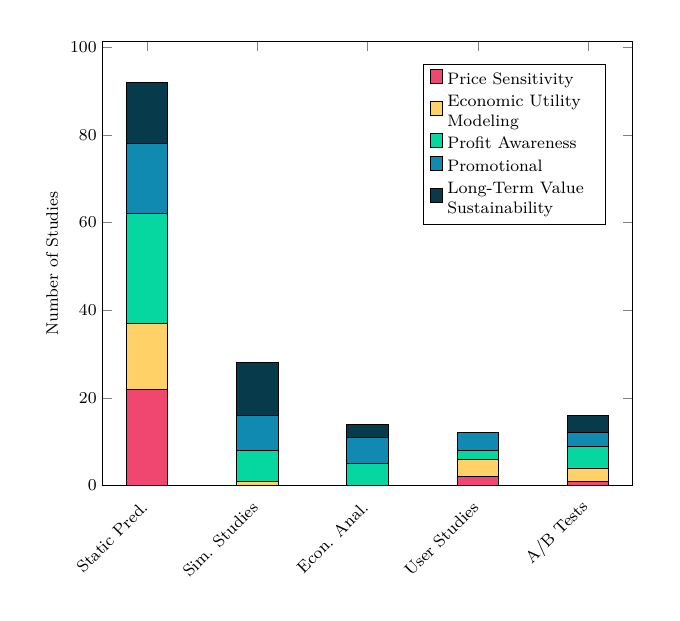}
\end{center}
%\vspace{-0.3cm}
\caption{Distribution of evaluation methods in the surveyed literature organized by dimension of analysis. %It might be helpful to print the content of the DAs here as well.
}
\label{fig:eval_method_distr}
% \vspace{-0.5cm}
\end{figure}

In the field of RSs, several methods are proposed to evaluate the performance of algorithms and systems.
Depending on the objective of the study, the evaluation may vary in order to assess specific aspects of the recommendations and the system.
We identify five methods that are used in the surveyed literature.
Some of these are used for offline evaluation (e.g., static predictions, simulation studies, and econometric analyses) \cite{zhao_revisiting_2023}, while others are used for online evaluation (e.g., user studies, and A/B tests) \cite{chen_performance_2017}.
While offline methods aim to give a plausible estimate of the performance the system could achieve under real circumstances if certain assumptions are verified, online ones are instead based on real user interactions.
In Figure~\ref{fig:eval_method_distr} we report the distribution of evaluation methods in the literature according to each DA.
As can be seen, offline methods are used more frequently than online ones.
Moreover, among offline methods, static predictions is the most frequently used method.

\paragraph{Static Predictions}

The most commonly used evaluation method in the RSs literature is to hide some data (e.g., ratings, interactions) from a particular dataset, train a model on the remaining data, and then predict the hidden data \cite{jannach_measuring_2019, zhao_revisiting_2023}.
After constructing a dataset that contains all the necessary information, the adopted standard is to measure the performance of the system with respect to some underlying objectives \cite{jannach_recommendations_2016, alslaity_goal_2021} with the help of certain metrics.
In terms of metrics, given the underlying purposes of ECRSs, the surveyed literature often measures not only relevance prediction metrics (e.g., precision, MRR, NDCG) \cite{ricci_evaluating_2022},
but also business value metrics\footnote{We discuss the most frequently used offline metrics in Section~\ref{sec:offline_metrics}.} (e.g., profit, revenue) \cite{jannach_measuring_2019, jannach_price_2017, malthouse_multistakeholder_2019}.

\paragraph{Simulation Studies}

While static predictions methods \cite{zhao_revisiting_2023} are mainly used to obtain an estimate of RSs performance in the short term,
other studies propose to use dynamic simulations to assess long-term performance \cite{ghanem_balancing_2022}.
The methodology first involves building a simulator to mimic user behavior \cite{pei_value-aware_2019, theocharous_personalized_2015, chbeir_click_2022}.
Next, the simulator is used to train and test RSs algorithms on the simulated behavior \cite{guo_maximizing_2020, wu_returning_2017}.
Simulators are often adopted to evaluate the performance of reinforcement learning-based recommendation algorithms \cite{afsar_reinforcement_2022} (e.g., RecoGym \cite{rohde_recogym_2018}, RecSim \cite{ie_recsim_2019}).
Moreover, in the surveyed literature there are also simulators \cite{iwata_recommendation_2008, lu_show_2014, ghanem_balancing_2022} created to evaluate supervised learning algorithms.
One of these \cite{ghanem_balancing_2022}, based on agent-based modeling, is designed to realistically mimic customer behavior considering various factors known in the literature to have a high correlation with purchase probability (e.g., trust).

\paragraph{Econometric Analyses}

For some algorithms in the surveyed literature \cite{das_maximizing_2009, hosanagar_recommended_2008, jiang_redesigning_2015}, performance is assessed with the help of  econometric analyses \cite{adamopoulos_business_2015, oestreicher-singer_visible_2012}.
These are quantitative approaches based on statistical or mathematical methods used to estimate the impact of the system on certain variables of interest (e.g., profit \cite{hosanagar_recommended_2008}), considering some underlying assumptions.
For example, one study \cite{hosanagar_recommended_2008} investigates the impact of recommendations on corporate profit and consumer welfare by modeling the behavior of a system that considers the simplified case in which the company can sell only two products.

\paragraph{User Studies}

In many cases, the impact of the system on certain factors (e.g., user satisfaction) is difficult to model through offline methods.
This occurs because in some cases it is not possible to find a good proxy for the target variable, while in other cases it would be necessary to use a large number of assumptions.
Especially when the factors are qualitative and the response is subjective (e.g., perceived fairness),
the literature adopts user studies as a research methodology \cite{chen_performance_2017}.
These methods typically involve recruiting a group of users (e.g., through emails or through crowdsourcing platforms like Amazon Mechanical Turk),
randomly splitting them into distinct groups,
requiring them to perform a particular task (e.g., interacting with an RS designed for the study),
observing their (objective) behavior, and asking them about their subjective perceptions.
In the surveyed literature these methods are used \cite{panniello_impact_2016, azaria_movie_2013} for example to determine the impact of algorithms on profitability and user trust.

\paragraph{A/B Tests}

When it is necessary to measure the performance of RSs in real-world circumstances, A/B tests are often performed \cite{chen_performance_2017}.
In such tests, two (or more) versions of a system are deployed for a certain period of time and users either interact with one or the other version \cite{jannach_measuring_2019}.
Although these tests are often complex to execute and require significant effort,
the main advantage is that they are able to directly measure business KPIs (e.g., revenue, profit) \cite{chen_performance_2017} and to compare different algorithms in production.
These tests are used many times in the surveyed literature\footnote{We discuss results of A/B tests in Section~\ref{sec:online_studies}.} since algorithms are often designed to optimize such KPIs.
For example, A/B tests are used to measure the effects of a profit-aware algorithm deployed on Alibaba's AliOS appstore \cite{zhang_smart_2017} and the CTR of a reinforcement learning-based algorithm deployed on a large e-commerce platform \cite{pei_value-aware_2019}.

\subsection{Metrics Used in Offline Evaluations}
\label{sec:offline_metrics}

\afterpage{\begingroup
\footnotesize
\begin{longtable}{m{0.5cm} m{7cm} m{1.2cm} m{5cm}} %[tp]

    \caption{Most frequently used offline evaluation metrics in the surveyed literature. *Note that for the sake of notation we used $p_j$ and $v_j$ as variables to indicate the price and profit of the recommended item at position $j$, but these variables depend only on the item and not on the position. **The formulas capture the main essence of the metrics.}
	\label{tab:metrics} \\
 \endlastfoot
        \hline
        Refs & Metric & Type & Definition \\
        \hline
\cite{ricci_evaluating_2022} & \vbox{\begin{equation}
        Prec@k = \frac{1}{|\mathcal{U}|}  \sum_{u \in \mathcal{U}} \frac{\sum_{j=1}^k{rel_{u,j}^{\mathcal{Y}}}}{k}
        \label{eq:precision}
    \end{equation} \vspace{-0.8cm}}                 & Relevance                 &  \textit{Precision} at position $k$ is the number of relevant items in the top-$k$ recommendations over the number of recommended ones. \\
\cite{ricci_evaluating_2022} & \vbox{\begin{equation}
        Rec@k = \frac{1}{|\mathcal{U}|}  \sum_{u \in \mathcal{U}} \frac{\sum_{j=1}^k{rel_{u,j}^{\mathcal{Y}}}}{\sum_{i=1}^n{x_{u,i}}}
        \label{eq:recall}
    \end{equation} \vspace{-0.7cm}}                 & Relevance                 &  \textit{Recall} at position $k$ is the number of relevant items in the top-$k$ recommendations over the total number of relevant ones. \\
\cite{ricci_evaluating_2022} & \vbox{\begin{equation}
        HR@k = \frac{1}{|\mathcal{U}|}  \sum_{u \in \mathcal{U}} \begin{cases}
        1 & \text{if } \quad \sum_{j=1}^k{rel_{u,j}^{\mathcal{Y}}} \ge 1 \\
        0 & \text{otherwise}
        \end{cases}
        \label{eq:hr}
    \end{equation} \vspace{-0.8cm}}                          & Relevance                 &  \textit{Hit-Rate} at position $k$ is the fraction of users for which the recommendations list contains at least one relevant item. \\
\cite{ricci_evaluating_2022} & \vbox{\begin{equation}
        MRR@k = \frac{1}{|\mathcal{U}|}  \sum_{u \in \mathcal{U}} \frac{1}{j_u^{1}}
        \label{eq:mrr}
    \end{equation} \vspace{-0.8cm}}                  & Relevance                 &  \textit{Mean Reciprocal Rank} at position $k$ is the mean rank of the first relevant item in the recommendations list. In the equation, $j_u^{1}$ is the rank (position) of the first item relevant to user $u$. \vspace{0.2cm} \\
\cite{ricci_evaluating_2022} & \vbox{\begin{equation}
        NDCG@k = \frac{1}{|\mathcal{U}|}  \sum_{u \in \mathcal{U}} \frac{\sum_{j=1}^k{\frac{rel_{u,j}^{\mathcal{Y}}}{log_2(j+1)}}}{IDCG_u@k}
        \label{eq:ndcg}
    \end{equation} \vspace{-0.2cm}}                 & Relevance                 & \textit{Normalized Discounted Cumulative Gain} at position $k$ is the inverse log reward on all positions with relevant items among the top-$k$ recommended ones. In the equation, $IDCG_u@k$ is the \textit{Ideal Discounted Cumulative Gain} obtained by sorting all the items relevant to the user in descending order. \\
\hline

\endfirsthead
\multicolumn{4}{c}{{\tablename\ \thetable{} -- Continued from previous page}} \\
\hline
Refs & Metric & Type & Definition \\
\hline
\endhead

\cite{malthouse_multistakeholder_2019}* & \vbox{\begin{equation}
        Revenue@k = \sum_{u \in \mathcal{U}} \sum_{j=1}^k{rel_{u,j}^{\mathcal{Y}} \cdot p_j}
        \label{eq:revenue}
    \end{equation} \vspace{-0.6cm}}                          & Value             & \textit{Revenue} at position $k$ is the revenue from relevant items in the recommendations list. \\
\cite{jannach_price_2017}* & \vbox{\begin{equation}
        Profit@k = \sum_{u \in \mathcal{U}} \sum_{j=1}^k{rel_{u,j}^{\mathcal{Y}} \cdot v_j}
        \label{eq:profit}
    \end{equation} \vspace{-0.6cm}}                           & Value             & \textit{Profit} at position $k$ is the profit from relevant items in the recommendations list. \\
\cite{cai_trustworthy_2019}* & \vbox{\begin{equation}
        EP@k = \sum_{u \in \mathcal{U}} \sum_{j=1}^k{\hat{x}_{u,j}(\mathbf{\Theta}) \cdot v_j}
        \label{eq:ep}
        \end{equation} \vspace{-0.6cm}}                             & Value                     &  \textit{Expected Profit} at position $k$ is the statistical profit it is expected to achieve by the recommendations considering the expected user interest $\hat{x}_{u,j}(\mathbf{\Theta})$. $EP@k$ is referred to as statistical profit (compared with $Profit@k$ in Eq.~\ref{eq:profit}), because the probability that the user accepts the recommendations instead of the actual ground truth relevance information is considered. \\
\cite{kompan_exploring_2022}** & \vbox{\begin{equation}
        PAH@k = \frac{1}{|\mathcal{U}|} \cdot \frac{Profit@k}{Volume@k}
        \label{eq:PAH}
    \end{equation} \vspace{-0.6cm}}                    & Value                     &  \textit{Profit-at-Hit} at position $k$ is the average profit per user from relevant items in the recommendations list. \\
\cite{louca_joint_2019}* & \vbox{\begin{equation}
        P\text{-}NDCG@k = \frac{1}{|\mathcal{U}|}  \sum_{u \in \mathcal{U}} \frac{\sum_{j=1}^k{\frac{rel_{u,j}^{\mathcal{Y}} \cdot p_j}{log_2(j+1)}}}{P\text{-}IDCG_u@k}
        \label{eq:pndcg}
    \end{equation} \vspace{-0.6cm}}                 & Value                     & \textit{Price-Based Normalized Discounted Cumulative Gain} at position $k$ is defined as $NDCG@k$, where the gain is given by the items price. In the equation, $P\text{-}IDCG_u@k$ is the \textit{Price-Based Ideal Discounted Cumulative Gain} obtained by sorting the prices of all relevant items to the user in descending order. \\

\hline \\

\end{longtable}
\endgroup}

\begin{comment}
Economic recommender systems are evaluated offline through various metrics.
In particular, besides the well-known prediction relevance metrics
 \cite{ricci_evaluating_2022}, additional ones are often combined to measure business value \cite{gotthardt_measuring_2022}.
In fact, considering especially the case of organization-oriented approaches (Sections~\ref{sec:profit_aware},~\ref{sec:promotion},~\ref{sec:long_term}), algorithms are often designed to recommend items of highest value for the organization rather than those of greatest interest for the user.
Hence, although \adrev{prediction relevance} metrics are still used to investigate
specific user-related aspects,
value metrics are used instead to investigate
organizational ones.
\end{comment}

A variety of metrics are used in the literature in offline evaluations, including both accuracy metrics to assess the relevance prediction performance as well as metrics aimed to investigate organizational value.

Given $rel_{u,j}^{\mathcal{Y}}$ as a ground truth relevance\footnote{The relevance of each item typically corresponds to the value of the ground truth $x_{u,i} \in \{ 0,1 \}$, i.e., assuming $x_{u,i} = 1$ if the item was actually purchased by the user, and $x_{u,i} = 0$ if not.} variable that indicates whether the item recommended at position $j$ in the ordered ranking $\mathcal{Y}_{u,k}$ is relevant or not for user $u$, we report in Table~\ref{tab:metrics} the metrics used for offline evaluations in the literature.
In the table we indicate for each metric the reference, the formula, the type, as well as its definition.
In the following, we mainly focus on value metrics, since relevance prediction metrics (e.g., Precision, Hit-Rate, NDCG - see Eq.~\ref{eq:precision},~\ref{eq:hr},~\ref{eq:ndcg}), are already widely  known \cite{ricci_evaluating_2022} and do not require further discussion here.

The general principle is the same for all value metrics.
Similarly to relevance prediction metrics, first a list of top-$k$ recommendations is generated for each user.
Then the recommendations are compared to the ground truth and certain value-related aspects are collected.
Those value-related aspects are connected to the price and profit (and in more general terms also to the utility)
of each recommended item.
In particular, differently from prediction relevance metrics, value ones do not only count the hits but multiply that hit by the items' price and profit.
We briefly introduce the most frequently used value metrics as follows:
\begin{itemize}
    \item \textit{Revenue@k} (Eq.~\ref{eq:revenue}) \cite{malthouse_algorithm_2019, malthouse_multistakeholder_2019, azaria_movie_2013} indicates the total revenue from the sale of recommended products actually purchased by users;
    \item \textit{Profit@k} (Eq.~\ref{eq:profit}) \cite{jannach_price_2017, ghanem_balancing_2022, concha-carrasco_multi-objective_2023, nemati_devising_2020} indicates the total profit from the sale of recommended products actually purchased by users;
    \item \textit{EP@k} (Eq.~\ref{eq:ep}) \cite{cai_trustworthy_2019} indicates the \textit{statistical} expected profit from the recommendation. \textit{EP@k} compared to \textit{Profit@k} is referred to as statistical because the probability of the user accepting the recommendations is considered rather than the ground truth information;
    \item \textit{PAH@k} (Eq.~\ref{eq:PAH}) \cite{kompan_exploring_2022} indicates the overall profit generated by the recommendation per user divided by the number of items sold;
    \item \textit{P-NDCG@k} (Eq.~\ref{eq:pndcg}) \cite{louca_joint_2019, lin_pareto-efficient_2019} indicates the total revenue generated on average per user from the recommendation compared to the theoretically achievable maximum revenue. \textit{P-NDCG@k}, like \textit{NDCG@k} (Eq.~\ref{eq:ndcg}) \cite{ricci_evaluating_2022}, gives more importance to the higher-priced items positioned on the top of the ranking\footnote{Note that, as in IR \cite{kobayashi_information_2000, said_information_2019}, value metrics can be rank-agnostic (e.g., \textit{Revenue@k}, \textit{Profit@k}) or rank-aware (e.g. \textit{P-NDCG@k}), depending on whether the position of the recommended items in the ranking is considered for evaluation or not.}.
\end{itemize}

However, analyzing the surveyed articles, some open issues can be identified.
In particular, we observe that the literature is mostly scattered, application-specific, and there are no well-defined standards in offline assessment of business value \cite{jannach_price_2017, debiasio_value_2023}.
Often the same metric is referred to by different names (e.g., \textit{Price-Based NDCG} \cite{louca_joint_2019}, vs.~\textit{G-DCG} \cite{lin_pareto-efficient_2019}).
Other times, researchers report results that are not comparable to each other because application-specific metrics are proposed in the article to investigate certain types of value (e.g., perishability \cite{seymen_making_2022}, marginal utility per dollar \cite{ge_maximizing_2019}).
In fact, under certain circumstances, it would not even be possible to use certain metrics.
For example, in the case where the underlying dataset carries only price information and not profit information (e.g., Amazon \cite{ni_justifying_2019}, Tmall \cite{zhu2018learning}), the metrics related to the latter would not be computable without using synthetic profit distributions of the dataset\footnote{We discuss the synthetic profit issue in Section~\ref{sec:dataset}.}.
Finally, in cases where simulations are used, the calculation of value metrics may be based on assumptions.
The main assumption that can be found \cite{jannach_price_2017, ghanem_balancing_2022} is that in some studies the user is supposed to always buy at least one item of the top-$k$ recommended ones.
In these cases, since the user may not have actually purchased any of the recommended items if his or her purchase history is analyzed, the underlying ground truth information may be unrealistic.

\subsection{Real-World A/B Tests and User Studies}
\label{sec:online_studies}

\begin{table*}[tp]
\fontsize{7}{7}\selectfont{
	\begin{center}
    \begin{tabular*}{\linewidth}{m{0.4cm} m{0.4cm}
    m{0.5cm}
    m{1.6cm} m{0.9cm} m{0.6cm}
    m{1.2cm} m{0.9cm}m{1cm}m{0.9cm}m{0.9cm}m{0.9cm}m{0.9cm} }
    \toprule
    Ref & Year & Eval.
    & Channel & Subjects & Durat.
    & Baseline & $\Delta\%$IPV & $\Delta\%$CTR & $\Delta\%$CVR & $\Delta\%$GMV & $\Delta\%$Rev. & $\Delta\%$Prof. \\
    \midrule

    \rowcolor[HTML]{D9D9D9}
\cite{maragheh_prospect-net_2022} & \citeyear{maragheh_prospect-net_2022} & A/B Test
& E-Commerce Platform (Walmart) & 36M sessions & -
& Walmart Ranker   &                                              &                                                &                                              & +\numprint{0.71}\%                           &                                             &                                                  \\
\cite{cavenaghi_online_2022}      & \citeyear{cavenaghi_online_2022} & A/B Test
& Booking Platform              & 1M searches  & 20 days
& Platform Ranker  &                                              & [-\numprint{0.50}\%, +\numprint{2.00}\%] &                                              &                                                &                                             &                                                  \\
\rowcolor[HTML]{D9D9D9}
\cite{agarwal_improving_2022}     & \citeyear{agarwal_improving_2022}  & A/B Test
& E-Commerce Platform           & -            & -
& Co-Purchase &                                              &                                                &                                              &                                                & +\numprint{35.0}\%                        &                                                  \\
\cite{li_revman_2021}             & \citeyear{li_revman_2021} & A/B Test
& Online Insurance Platform     & -            & 1 week
& LogReg          &                                              &                                                & +[\numprint{1.05}\%, \numprint{3.98}\%] &                                                & +[\numprint{2.7}\%, \numprint{16.2}\%] &                                                  \\
\rowcolor[HTML]{D9D9D9}
\cite{ji_reinforcement_2021}      & \citeyear{ji_reinforcement_2021}  & A/B Test
& E-Commerce Platform (Taobao)  & -            & 1 week
& Vanilla-CTR                  & +[\numprint{6.25}\%, \numprint{8.67}\%] &                                                &                                              & +[\numprint{12.31}\%, \numprint{18.03}\%] &                                             &                                                  \\
\cite{deng_personalized_2020}     & \citeyear{deng_personalized_2020}  & A/B Test
& Video Game Platform (NetEase) & -            & 1 year
& Platform Ranker  &                                              &                                                & +\numprint{60.0}\%                         & +\numprint{15.0}\%                           &                                             &                                                  \\
\rowcolor[HTML]{D9D9D9}
\cite{pei_value-aware_2019}       & \citeyear{pei_value-aware_2019}  & A/B Test
& E-Commerce Platform           & 1M users     & 1 week
& Item KNN                    & +\numprint{8.80}\%                         & +\numprint{8.20}\%                           &                                              & +\numprint{27.90}\%                          &                                             &                                                  \\
\cite{lin_pareto-efficient_2019}  & \citeyear{lin_pareto-efficient_2019}  & A/B Test
& E-Commerce Platform           & -            & 3 days
& LETORIF                      & +\numprint{23.76}\%                        & +\numprint{13.80}\%                          &                                              & +\numprint{3.62}\%                           &                                             &                                                  \\
\rowcolor[HTML]{D9D9D9}
\cite{zhang_smart_2017}           & \citeyear{zhang_smart_2017}  & A/B Test
& Appstore (Alibaba AliOS)      & 1M users     & 2 weeks
& LinDP                        &                                              &                                                & -\numprint{6.00}\%                         &                                                &                                             & +\numprint{32.0}\%                             \\
\midrule
\cite{panniello_impact_2016}*      & \citeyear{panniello_impact_2016}  & User Study
& Mail Campaign                 & 260 users    & 9 weeks
& CBF      &                                              &                                                &                                              &                                                & +\numprint{94.39}\%                        & +\numprint{137}\%                           \\
\rowcolor[HTML]{D9D9D9}
\cite{zhao_e-commerce_2015}       & \citeyear{zhao_e-commerce_2015}  & User Study
& Amazon Mechanical Turk        & 79 users     & -
& Amazon Price                 &                                              &                                                &                                              &                                                &                                             & +[\numprint{241}\%, \numprint{248}\%] \\
\cite{zhu_bundle_2014}            & \citeyear{zhu_bundle_2014}    & User Study
& Mail Campaign
& few users & 1 week
& Markov Model                 &                                              & +\numprint{7.43}\%                           & +\numprint{48.92}\%                        &                                                &                                             &                                                  \\
\rowcolor[HTML]{D9D9D9}
\cite{azaria_movie_2013}          & \citeyear{azaria_movie_2013}  & User Study
& Amazon Mechanical Turk        & 245 users    & -
& Pers. NonCF &                                              &                                                &                                              &                                                & +\numprint{28.57}\%                       &                                                 \\

    \bottomrule
    \end{tabular*}
    \end{center}
}
\vspace{-0.6cm}
    \caption{Results of real-world A/B tests and user studies in the surveyed literature. In the table we report:
    the channel used to convey the recommendations;
    the number of subjects (i.e., users, searches, or sessions);
    the overall duration of the study (e.g. 20 days, 1 year);
        the baseline against which the proposed algorithm in the study is compared;
    and the relative improvements in online metrics of the proposed algorithm compared to the baseline. *The relative improvements
    are determined by analyzing the sentence \quotes{\textit{Overall revenue generated during the experiment was €428 for the content-based group, €832 for the profit-based group}} and Figure 11b in the original paper \cite{panniello_impact_2016}.}
    \vspace{-0.5cm}
	\label{tab:online_studies}
\end{table*}

\begin{table}[tp]
\footnotesize
	\begin{center}
    \begin{tabular}{m{1cm} m{14cm} }
    \toprule
    Metric & Meaning \\
    \midrule
    IPV & \textit{Individual Page View} is the overall number of clicked items on the platform. \\
    CTR & \textit{Click-Through Rate} is the number of user clicks divided by the number of items shown. \\
    CVR & \textit{Click-Conversion Rate} is the number of purchases (or other events) divided by the number of clicks.\\
    GMV* & \textit{Gross Merchandise Value} is the number of items sold multiplied by their price.\\
    Revenue* & \textit{Revenue} is equal to GMV minus any commissions from item sellers.\\
    Profit** & \textit{Profit} is equal to Revenue minus any item costs.\\
    \bottomrule
    \end{tabular}
    \end{center}
    \vspace{-0.3cm}
    \caption{Most frequently used online metrics in the surveyed literature \cite{jannach_measuring_2019}. *GMV and Revenue almost always indicate the same measure except in B2C marketplaces like eBay. **Depending on the type of subtracted cost (e.g., raw materials, marketing), profit can be gross, net, or have additional nuances (e.g., EBITDA, EBIT).
    \vspace{-0.1cm}}
	\label{tab:online_metrics}
\end{table}

Many authors evaluate the performance of ECRSs algorithms using A/B tests or user studies.
As is known in the literature, offline evaluation results are not necessarily a valid indicator of online performance \cite{jannach_measuring_2019, jannach_case_2009}.
This is often due to the fact that different metrics are used for the two types of experimental evaluation \cite{dias_value_2008, senecal_influence_2004}.
While offline metrics are often used to measure relevance prediction accuracy (e.g., Precision, NDCG), online metrics are used instead to measure business value (e.g., CTR, GMV, Revenue) \cite{hutchison_live_2012, garcin_offline_2014}.
Companies are usually much more interested in assessing how algorithms impact real-world business KPIs exploiting online metrics.

In Table~\ref{tab:online_studies} we list the studies in the surveyed literature that measure the performance of the proposed systems through A/B tests or user studies.
In Table~\ref{tab:online_metrics} we then briefly summarize the meaning of each online metric that is considered for the analysis\footnote{Some niche metrics used to measure certain application-specific factors reported in the studies are not considered} (i.e., IPV, CTR, CVR, GMV, Revenue, Profit).
We refer readers to a recent survey \cite{jannach_measuring_2019} on this topic for further insights into online metrics.

Analyzing Table~\ref{tab:online_studies} we can make some interesting observations.
Some considerations depend on the nature of the particular evaluation methodology (i.e., A/B test vs.~user study).
For example, considering the recommendations channel and the number of subjects, we note that user studies typically involve few users recruited through e-mail campaigns \cite{panniello_impact_2016, zhu_bundle_2014} or Amazon Mechanical Turk \cite{azaria_movie_2013, zhao_e-commerce_2015}.
Instead, A/B tests are typically performed on a large scale, exploiting existing systems with large customer bases \cite{zhang_smart_2017, pei_value-aware_2019}, some of well-known brands (e.g., Walmart, Taobao, Alibaba, NetEase) \cite{maragheh_prospect-net_2022, zhang_smart_2017, deng_personalized_2020, ji_reinforcement_2021}.
Moreover, from a performance point of view, all the studies, whether they are based on user studies or A/B tests, show that ECRSs are able to potentially bring huge business value to the firm.
In fact, increases in online metrics are reported in all studies.
In some cases, the authors report significant performance improvements\footnote{To ensure evaluation reliability, many authors test the proposed algorithm in different configuration environments reporting different results for each of them \cite{li_revman_2021}.
In these cases, Table~\ref{tab:online_studies} shows a range instead of a single value in metrics improvement.} (e.g., +$48.92\%$ CVR \cite{zhu_bundle_2014}, +$35\%$ revenue \cite{agarwal_improving_2022}, +$32\%$ profit \cite{zhang_smart_2017}).

However, there may be some limitations regarding the insights we can get from the studies.
For example, most of the A/B tests last a very short time, i.e., less than three weeks\footnote{Performing long-term A/B tests on a real platform is complex \cite{jannach_measuring_2019} and significant effort is required both in the planning and analysis phases.
Often the test could cause financial damage to the brand as users could lose trust in the company due to ineffective recommendations.
Other times, it is necessary to re-run the test because of bugs.
Or again, certain events (e.g., Easter, Super Bowl) or global macroenomic circumstances (e.g., 2020 COVID-19 crisis, 2022 Ukranian war) may impact performance.} \cite{cavenaghi_online_2022, li_revman_2021, ji_reinforcement_2021, pei_value-aware_2019, lin_pareto-efficient_2019, zhang_smart_2017, zhao_e-commerce_2015, zhu_bundle_2014, azaria_movie_2013}.
In some cases, the baselines are proprietary algorithms and their internal mechanisms are unknown \cite{maragheh_prospect-net_2022, cavenaghi_online_2022, deng_personalized_2020, azaria_movie_2013} (e.g., Walmart Ranker).
In other cases, results depend on assumptions.
For example, a study \cite{zhao_e-commerce_2015} based on Amazon Mechanical Turk uses synthetic profit information, as the authors did not have product costs available.
Another study \cite{panniello_impact_2016} uses some proxies for offline purchases in addition to explicit purchase data from the firm’s online site to measure revenue.
In that specific context, offline purchases cannot be connected to the online identities of experiment participants.
In particular, the authors treated items that received high ratings by users after they clicked on the \quotes{\textit{see more details}} link as purchases to calculate profit.

\subsection{Available Datasets}
\label{sec:dataset}

Analyzing the ECRSs literature, our survey reveals that many studies report results based on proprietary datasets.
This is mainly due to the fact that certain types of information (e.g., prices, profits, purchases, demographics) are of strategic importance to companies, and uncontrolled sharing could create significant economic damage.
For example, some information is sensitive to the user, and non-anonymized sharing could have major legal implications due to privacy laws, as well as significant impact on  brand reputation.
In addition, competitors could make use of economic data related to purchasing and profitability to study weaknesses in the business model and take away market share.
However, especially recently, several studies also report results based on public datasets.

\begin{table*}[tp]
\fontsize{8}{8}\selectfont
	\begin{center}
    \begin{tabular*}{\linewidth}{m{0.4cm} l m{1.2cm} m{1.2cm}  m{1.2cm}  m{1.3cm} m{1.7cm}m{0.4cm}m{0.4cm}m{0.4cm}m{0.4cm}m{0.4cm} }
    \toprule
    Ref & Dataset & \#User & \#Item & \#Inter & Density & Event & Date & Dem. & Cat. & Price & Prof.\\
    \midrule
\rowcolor[HTML]{D9D9D9}
\cite{kaggle2019commerce}                 & \href{https://www.kaggle.com/datasets/mkechinov/ecommerce-events-history-in-cosmetics-shop}{Cosmetics} & \num{1639358}  & \num{54571}     & \num{20692840}  & \numprint{0.023} \%           & View,  Add-to-Cart, Remove-From-Cart, Purchase & \checkmark &            & \checkmark & \checkmark &            \\
\cite{cup2016track}                       & \href{https://competitions.codalab.org/competitions/11161}{Diginetica}                                        & \num{204789}   & \num{184047}    & \num{993483}    & \numprint{0.002} \%           & Query, Click, Purchase                        & \checkmark &            & \checkmark & \checkmark &            \\
\rowcolor[HTML]{D9D9D9}
\cite{ni_justifying_2019}                 & \href{https://nijianmo.github.io/amazon/index.html}{Amazon(2018)*}                                            & -              & \num{15500000}  & \num{233000000} & -                             & Review, Ratings                               & \checkmark &            & \checkmark & \checkmark &            \\
\cite{yelp2016dataset}                    & \href{https://www.yelp.com/dataset}{Yelp(Full)*}                                                              & \num{5556436}  & \num{539254}    & \num{289088240} & \numprint{0.009} \%           & Review, Ratings                               & \checkmark &            & \checkmark & \checkmark &            \\
\rowcolor[HTML]{D9D9D9}
\cite{dror2012yahoo}                       & \href{https://webscope.sandbox.yahoo.com/catalog.php?datatype=r}{Yahoo!Music}                                 & \num{1948882}  & \num{98211}     & \num{11557943}  & \numprint{0.006} \%           & Ratings                                       &            &            &            &            &            \\
\cite{hsu2004mining}                      & \href{https://www.kaggle.com/datasets/chiranjivdas09/ta-feng-grocery-dataset}{Ta-Feng}                        & \num{32266}    & \num{23812}     & \num{817741}    & \numprint{0.106} \%           & Purchase                                      & \checkmark & \checkmark & \checkmark & \checkmark &            \\
\rowcolor[HTML]{D9D9D9}
\cite{harper_movielens_2016}              & \href{https://grouplens.org/datasets/movielens/}{MovieLens(20M)*}                                             & \num{138493}   & \num{27278}     & \num{20000263}  & \numprint{0.529} \%           & Ratings                                       & \checkmark & \checkmark & \checkmark &            &            \\
\cite{bennett2007netflix}                 & \href{https://www.kaggle.com/datasets/netflix-inc/netflix-prize-data}{Netflix Prize}                          & \num{480189}   & \num{17770}     & \num{100480507} & \numprint{1.177} \%           & Ratings                                       & \checkmark &            &            &            &            \\
\rowcolor[HTML]{D9D9D9}
\cite{berendt_spmf_2016}                  & \href{https://www.philippe-fournier-viger.com/spmf/index.php?link=datasets.php}{SPMF}                         & -              & \num{16470}     & \num{88162}     & -                             & Purchase                                      & \checkmark &            &            & \checkmark &            \\
\cite{lin_pareto-efficient_2019}          & \href{https://drive.google.com/open?id=1rbidQksa_mLQz-V1d2X43WuUQQVa7P8H}{EC-REC}                             & -              & -               & -               & -                             & View, Click, Purchase                         & \checkmark &            &            & \checkmark &            \\
\rowcolor[HTML]{D9D9D9}
\cite{ziegler_improving_2005}             & \href{http://www2.informatik.uni-freiburg.de/~cziegler/BX/}{Book-Crossing}                                    & \num{105284}   & \num{340557}    & \num{1149780}   & \numprint{0.003} \%           & Ratings                                       &            & \checkmark &            &            &            \\
\cite{microsoft_corporation_example_1998} & \href{https://github.com/julianhyde/foodmart-data-hsqldb}{Foodmart}                                           & \num{8842}     & \num{1559}      & \num{261130}    & \numprint{1.894} \%           & Purchase                                      & \checkmark & \checkmark & \checkmark & \checkmark & \checkmark \\
\rowcolor[HTML]{D9D9D9}
\cite{mcfee2012million}                   & \href{http://millionsongdataset.com/lastfm/}{Last.fm}                                                         & \num{1892}     & \num{176322}    & \num{92834}     & \numprint{0.027} \%           & Listen                                        & \checkmark &            & \checkmark &            &            \\
\cite{richardson_mining_2002}             & \href{https://snap.stanford.edu/data/soc-Epinions1.html}{Epinions}                                            & \num{226570}   & \num{231637}    & \num{1132367}   & \numprint{0.002} \%           & Ratings, Graph                                & \checkmark &            &            &            &            \\
\rowcolor[HTML]{D9D9D9}
\cite{goldberg_eigentaste_2001}           & \href{https://eigentaste.berkeley.edu/dataset/}{Jester}                                                       & \num{73421}    & \num{101}       & \num{4136360}   & \numprint{55.779} \%          & Ratings                                       &            &            &            &            &            \\
\cite{pathak2017generating}               & \href{https://github.com/kang205/SASRec}{Steam}                                                               & \num{2567538}  & \num{32135}     & \num{7793069}   & \numprint{0.009} \%           & Purchase                                      & \checkmark &            & \checkmark & \checkmark &            \\
\rowcolor[HTML]{D9D9D9}
\cite{zhu2018learning}                    & \href{https://tianchi.aliyun.com/dataset/649}{Tmall}                                                          & \num{963923}   & \num{2353207}   & \num{44528127}  & \numprint{0.001} \%           & View,  Add-to-Cart, Add-to-Wishlist, Purchase  & \checkmark &            & \checkmark &            &            \\
\cite{pei_value-aware_2019}               & \href{https://github.com/rec-agent/rec-rl}{REC-RL}                                                            & \num{49000000} & \num{200000000} & \num{763000000} & \num{7.78571428571429E-06} \% & Click,  Add-to-Cart, Add-to-Wishlist, Purchase & \checkmark &            &            & \checkmark &            \\
\rowcolor[HTML]{D9D9D9}
\cite{ventatesan_dunnhumby_2007}          & \href{https://www.kaggle.com/datasets/frtgnn/dunnhumby-the-complete-journey}{Dunnhumby}                       & \num{2500}     & \num{92400}     & \num{2600000}   & \numprint{1.125} \%           & Purchase                                      & \checkmark & \checkmark & \checkmark & \checkmark &            \\
\cite{pisharath_nu-minebench_2005}        & \href{http://cucis.ece.northwestern.edu/projects/DMS/MineBench.html}{Chainstore}                              & -              & \num{46086}     & \num{1112949}   & -                             & Purchase                                      &            &            &            & \checkmark &            \\
\rowcolor[HTML]{D9D9D9}
\cite{dvn_tgwx_2018}                      & \href{https://github.com/cjx0525/BGCN}{NetEase}                                                               & \num{18528}    & \num{123628}    & \num{1128065}   & \numprint{0.049} \%           & Playlist                                      &            &            &            &            &           \\

    \bottomrule
    \end{tabular*}
    \end{center}
    \vspace{-0.3cm}
    \caption{Most used datasets in the surveyed literature. *Since there may be multiple versions of the same dataset, we report the statistics of the most recent one.}
    \vspace{-0.5cm}
	\label{tab:datasets}
\end{table*}

In Table~\ref{tab:datasets} we report the most frequently used public datasets in the surveyed literature.
Specifically, in addition to statistical information such as the number of users, items, interactions, and the density of the dataset, we also report the type of event/interaction (e.g., click, add-to-cart, purchase, rating), and the presence of relevant features for ECRSs algorithms, i.e., date, user demographics, product category, price, and profit.

Analyzing the reported information, we can make some observations.
First, both the datasets' density and size, i.e., the number of interactions, vary greatly.
Some of them are quite sparse (e.g., REC-RL \cite{pei_value-aware_2019}), whereas others are dense (e.g., Jester \cite{goldberg_eigentaste_2001}).
Some are quite small (e.g., Foodmart \cite{microsoft_corporation_example_1998}), while others are large (e.g., Amazon \cite{ni_justifying_2019}).
In addition, as expected, most of the datasets contain economic information related to actual purchases, as well as prices and possibly profit of products (e.g., in the Cosmetics \cite{kaggle2019commerce}, Diginetica \cite{cup2016track}, Ta-Feng \cite{hsu2004mining}, and Tmall \cite{zhu2018learning} datasets).
Indeed, as discussed earlier, economic information is typically used for both algorithmic and evaluation purposes.
However, as can be noted, some datasets do not contain prices (e.g., MovieLens \cite{harper_movielens_2016}, Netflix Prize \cite{bennett2007netflix}, Book-Crossing \cite{ziegler_improving_2005}, Epinions \cite{richardson_mining_2002}, Last.fm \cite{mcfee2012million}) and currently only Foodmart \cite{microsoft_corporation_example_1998} contains profit.
We observe that those datasets are the most frequently used in RSs research.
In particular, although profit is very important especially to train profit-aware models, we note various studies \cite{jannach_price_2017, ghanem_balancing_2022, brand_random_2005, akoglu_valuepick_2010, piton_capre_2011, lu_show_2014, cai_trustworthy_2019, nemati_devising_2020, concha-carrasco_multi-objective_2023} assuming some synthetic profit distribution, e.g., normal \cite{ghanem_balancing_2022}, or random \cite{nemati_devising_2020}.
This assumption would allow to overcome the profit availability issue.
However, as reported in almost all the studies, this also constitutes an important limitation.
In fact, under real circumstances, the profit distribution could be very different from the synthetic one used for the experiments, and the results could vary considerably.

\subsection{Reproducibility Maturity}

The impact of reproducibility on the progress of science is undeniable.
However, although there has generally been an increase in reproducible papers in AI over the years \cite{gundersen_state_2018}, many of them are still not sufficiently well documented to reproduce the results of the reported experiments \cite{haibe-kains_transparency_2020}.
This problem is observed several times in the field of RSs \cite{beel_towards_2016, cremonesi_progress_2021}, with well-known cases regarding articles that proposed neural algorithms \cite{rendle_neural_2020, ferrari_dacrema_are_2019}, highlighting for example:
non-uniform and lax standards in adopting the correct experimental evaluation methodologies \cite{sun_are_2020};
questionable choices on the use and fine-tuning of baselines for comparative experiments \cite{ferrari_dacrema_troubling_2021}.

\begin{table}[tp]
\footnotesize
\sloppy

	\begin{center}
    \begin{tabular}{m{0.4cm} m{0.4cm} m{2.5cm} m{10cm} }
    \toprule
    Ref & Year & Dimension & Link \\
    \midrule
\rowcolor[HTML]{D9D9D9}
\cite{chang_bundle_2023}         & \citeyear{chang_bundle_2023}         & Promotional
& \url{https://github.com/cjx0525/BGCN}                                                               \\
\cite{zhang_price_2022}          & \citeyear{zhang_price_2022}          & Price-Sensitivity    & \url{https://github.com/Zhang-xiaokun/CoHHN}                                                        \\
\rowcolor[HTML]{D9D9D9}
\cite{wu_cheaper_2022}           & \citeyear{wu_cheaper_2022}           & Price-Sensitivity    & \url{https://github.com/PCNet-Code}                                                                 \\
\cite{ghanem_balancing_2022}     & \citeyear{ghanem_balancing_2022}     & Profit-Awareness     & \url{https://github.com/nadaa/rec-strategies-abm}                                                   \\
\rowcolor[HTML]{D9D9D9}
\cite{avny_brosh_bruce_2022}     & \citeyear{avny_brosh_bruce_2022}     & Promotional
& \url{https://github.com/tzoof/BRUCE}                                                                \\
\cite{agarwal_improving_2022}    & \citeyear{agarwal_improving_2022}    & Promotional
& \url{https://github.com/muhanzhang/SEAL}                                                            \\
\rowcolor[HTML]{D9D9D9}
\cite{zhan_towards_2021}         & \citeyear{zhan_towards_2021}         & Long-Term Value Sustainability     & \url{https://github.com/google-research/google-research/tree/master/recs_ecosystem_creator_rl}       \\
\cite{zheng_price-aware_2020}    & \citeyear{zheng_price-aware_2020}    & Price-Sensitivity    & \url{https://github.com/DavyMorgan/ICDE20-PUP}                                                     \\
\rowcolor[HTML]{D9D9D9}
\cite{xu_e-commerce_2020}        & \citeyear{xu_e-commerce_2020}        & Economic Utility Modeling   & \url{https://github.com/zhichaoxu-shufe/E-commerce-Rec-with-WEU}                                    \\
\cite{ge_learning_2020}          & \citeyear{ge_learning_2020}          & Economic Utility Modeling    & \url{https://github.com/TobyGE/Risk-Aware-Recommnedation-Model}                                      \\
\rowcolor[HTML]{D9D9D9}
\cite{dai_u-rank_2020}           & \citeyear{dai_u-rank_2020}           & Economic Utility Modeling    & \url{https://github.com/xydaisjtu/U-rank}                                                          \\
\cite{chang_bundle_2020}         & \citeyear{chang_bundle_2020}         & Promotional
& \url{https://github.com/cjx0525/BGCN}                                                              \\
\rowcolor[HTML]{D9D9D9}
\cite{pei_value-aware_2019}      & \citeyear{pei_value-aware_2019}      & Long-Term Value  Sustainability    & \url{https://github.com/rec-agent/rec-rl}                                                          \\
\cite{lin_pareto-efficient_2019} & \citeyear{lin_pareto-efficient_2019} & Profit-Awareness     & \url{https://github.com/weberrr/PE-LTR}                                                             \\
\rowcolor[HTML]{D9D9D9}
\cite{ge_maximizing_2019}        & \citeyear{ge_maximizing_2019}        & Economic Utility Modeling     & \url{https://github.com/TobyGE/Maximizing-Marginal-Utility-per-Dollar-for-Economic-Recommendation} \\
    \bottomrule
    \end{tabular}
    \end{center}
    \vspace{-0.3cm}
    \caption{Studies in the surveyed literature that provide the code.}
    \vspace{-0.5cm}
	\label{tab:reproducibility}
\end{table}

In particular, by reviewing the ECRSs literature, we note several limitations concerning the reproducibility of the studies.
As reported in Table~\ref{tab:reproducibility}, only a very small subset of 15 articles, out of 135 ($11.11$\%) identified by the present systematic review share the implementation code\footnote{We did not dive into the code details because even if the code is shared, it was found earlier in the RSs literature \cite{sun_are_2020, ferrari_dacrema_troubling_2021, cremonesi_progress_2021} that in many cases important information is missing  to ensure reproducibility (e.g., pre-processing code).}.
Notably, as can be seen from the table, we find no article that publicly share the code prior to 2019.
In addition, the level of reproducibility is quite uneven when considering the different subdomains of ECRSs.
In particular, we note the following critical issues:
there are many articles published in the \textit{profit-awareness} subdomain but only two of them share the code;
all the articles published in the field of \textit{promotional} strategies refer to relevance-based bundling methods (i.e., there is no code shared about brand-awareness and pricing methods);
the code of articles concerning \textit{price-sensitivity} and \textit{long-term value} methods is published only for the most recent and advanced GNN- and RL-based algorithms.
Consequently, it would be beneficial and significantly accelerate progress in this field if researchers would pay special attention to increasing the level of reproducibility.

\section{Current Challenges and Future Research}
\label{sec:open_challenges}

In this section we discuss current challenges of ECRSs
and possible future research directions.

\paragraph{Comparing Different Algorithmic Approaches}

A multitude of algorithmic approaches for optimizing business value are proposed in the literature.
In this paper, we categorize them at a high level into in-processing and post-processing methods \cite{debiasio_value_2023} considering five DAs.
However, most of the approaches are never compared with each other and may have specificities that may make them preferable in certain circumstances over others.
For example, no study has yet compared in-processing with post-processing approaches.
In addition, different types of in-processing algorithms are found in the literature.
In particular,
it is proposed for example to extend the objective function of MF \cite{ge_cost-aware_2014, chen_does_2014, chen_boosting_2017, sato_discount_2015},
or to use GNNs \cite{zheng_price-aware_2020, zheng_incorporating_2021, zhang_price_2022} to generate price-sensitive recommendations.
Moreover, value neighbor selection \cite{cai_trustworthy_2019}, graph-based \cite{brand_random_2005, akoglu_valuepick_2010, qu_cost-effective_2014} or evolutionary \cite{nemati_devising_2020, concha-carrasco_multi-objective_2023} profit-aware algorithms are proposed as well.
However, some types of methods are applied only to certain DAs.
For example, although feasible in practice, no profit-aware MF objective function extensions or GNNs were surfaced through our study.
Similarly, no neighbor selection or evolutionary price-sensitive algorithm was found so far.
Therefore, it might be useful for the future both to compare in-processing and post-processing approaches
and to implement theoretically feasible algorithms not yet found in the literature, comparing them with existing ones.

\paragraph{Optimizing Business Value Trade-Offs}

Business value optimization is complex, and the systems must consider multiple trade-offs \cite{jannach_price_2017, debiasio_value_2023} in the optimization process.
For example, considering real-world businesses based on an advertising revenue model (e.g., YouTube, Alibaba's AliOS), it is very important to find the right balance between the ad revenue generated by sponsored items and the actual interests of the user \cite{zhang_smart_2017, malthouse_multistakeholder_2019}.
In particular, special care must be taken not to compromise user trust \cite{panniello_impact_2016, hosanagar_recommended_2008}.
In fact, it is shown both through simulations \cite{ghanem_balancing_2022}, in user studies \cite{nilashi_recommendation_2016} and subsequent A/B tests \cite{panniello_impact_2016} that trust is positively correlated with propensity to purchase.
A system that is too biased toward higher-value items that provides irrelevant recommendations to the user \cite{jannach_price_2017, debiasio_value_2023} could risk impacting the organization's reputation and driving away customers.
To address this issue various studies \cite{azaria_movie_2013, wang_mathematical_2009, lu_show_2014, malthouse_multistakeholder_2019, zhang_smart_2017, kompan_exploring_2022} propose algorithms with the goal of balancing the interests of multiple stakeholders \cite{abdollahpouri_multistakeholder_2020, ricci_multistakeholder_2022}, particularly considering the profitability/relevance trade-off \cite{jannach_price_2017}, and optimizing short- or long-term value \cite{hosein_recommendations_2019}.
Furthermore, as various studies pointed out, algorithms should take care also of explainability \cite{zhang_explainable_2020, tintarev_survey_2007, montagna2023graph}, fairness \cite{zehlike_fairness_2023, zehlike_fairness_2023_2, patro_fair_2022, pitoura_fairness_2022}, and diversity \cite{kunaver_diversity_2017, panniello_impact_2016} since they are directly related to trust \cite{deng_personalized_2020}.
However, the current literature has not thoroughly investigated the impact of many of these factors on business value.
Hence, providing efficient algorithms to simultaneously optimize multiple business value trade-offs (e.g., profit, fairness, and trust) could be a valuable research direction for the future.

\paragraph{Comprehensive Purpose-Oriented Offline and Online Evaluation}

Evaluating ECRSs often requires the use of methods that are different from those used for traditional RSs \cite{zhao_revisiting_2023, chen_performance_2017}.
As a result, there are still many open challenges in order to to be able to evaluate ECRSs in a comprehensive, purpose-oriented way \cite{jannach_recommendations_2016, alslaity_goal_2021} (i.e., that considers the purposes for which the system is designed).
Several of these challenges follow from the analysis presented in this paper.
For example, in offline evaluation, it is necessary to use business value metrics besides the widely adopted relevance prediction metrics \cite{ricci_evaluating_2022}.
Studies often exploit a variety of metrics \cite{guo_maximizing_2020, cai_trustworthy_2019, jannach_price_2017, kompan_exploring_2022, louca_joint_2019} albeit with similar objectives, and the results reported are not comparable with each other.
In addition, offline evaluation methodologies are not standardized and often are designed ad-hoc according to specific needs \cite{hosanagar_recommended_2008}.
Moreover, besides a few exceptions \cite{ge_maximizing_2019, ghanem_balancing_2022, zhang_price_2022, zhan_towards_2021, pei_value-aware_2019, xu_e-commerce_2020, lin_pareto-efficient_2019}, most studies are difficult to reproduce and are often based on proprietary datasets or public datasets with synthetic data \cite{nemati_devising_2020, concha-carrasco_multi-objective_2023}.
In fact, most datasets \cite{harper_movielens_2016, bennett2007netflix, kaggle2019commerce, ni_justifying_2019, hsu2004mining, richardson_mining_2002, ziegler_improving_2005} do not contain information such as profitability \cite{microsoft_corporation_example_1998}, which is however needed for model training.
Regarding A/B tests on the other hand, many of them last for a short time \cite{li_revman_2021, pei_value-aware_2019, lin_pareto-efficient_2019} and involve a small set of users \cite{zhu_bundle_2014, azaria_movie_2013, zhao_e-commerce_2015, panniello_impact_2016} to avoid potential economic risks \cite{jannach_measuring_2019, panniello_impact_2016} for the organization hosting the test.
Hence, there could be several future research directions in the field of evaluation.
For example, it is necessary to develop better offline value metrics that are indicative of online performance in a given (prototype) scenario.
In addition, large-scale A/B tests (i.e., involving many users) and reproducibility studies are also required.

\paragraph{Design of Holistic Algorithmic Methods}

In this work, decomposing the literature on ECRSs into five different DAs, various algorithmic approaches for optimizing business value are explored.
However, most of the existing methods \cite{chen_developing_2008, pei_value-aware_2019, chen_boosting_2017, ge_maximizing_2019, zhao_e-commerce_2015} focus exclusively on one of the five perspectives.
There are a few exceptions \cite{kompan_exploring_2022, maragheh_prospect-net_2022} that involve more than one DA that study, for example \cite{kompan_exploring_2022}, how to combine price-sensitivity with profit-awareness to generate more profit while keeping relevance high.
A very small subset of studies \cite{demirezen_optimization_2016, seymen_making_2022, ettl_data-driven_2020}, on the other hand, provide broader reasoning by also discussing inventory management techniques that might be useful for analogous purposes.
Currently, the literature lacks holistic methods capable of leveraging multiple approaches simultaneously \cite{debiasio_value_2023, jannach_price_2017} complementing each other to optimize different nuances of business value \cite{jannach_measuring_2019} while also considering the interrelationship \cite{he_propn_2022} between them.
In addition, it is also necessary to consider the relationship between sales and marketing processes with operational \cite{seymen_making_2022, demirezen_optimization_2016, ettl_data-driven_2020} and financial processes so as to propose methods for improving the entire business ecosystem, e.g., reducing raw material costs, minimizing logistics delays, or optimizing cash flows.

\section{Conclusion and Implications}
\label{sec:conclusion}

In this paper, we review the existing literature on ECRSs.
Unlike traditional RSs, economic ones aim to directly optimize profitability by exploiting purchase information (e.g., price and profit) and related concepts from economics and marketing.
This topic is highly important because organizations aim to optimize (long-term) profit.
Accordingly, ECRSs are well-suited for use in commercial applications such as e-commerces, media streaming sites, and advertising platforms, as they offer various benefits for organizations to increase their business KPIs.
In this survey, we identify a number of relevant works addressing a multitude of related issues on ECRSs.
In particular, although the literature is highly scattered, five different approaches that jointly consider the interests of customers and organizations are identified in this paper (e.g., price sensitivity, profit awareness).

At present, the application of a certain approach in any company can be viewed as a strategic management decision.
Indeed, while all approaches are useful (and can hypothetically be applied at the same time) for increasing corporate profitability, some may be more effective than others in a specific business context.
The management should make wise decisions about which approach to employ by considering, for example, the company's particular business strategy and revenue model, the expected business value returns and the potential behavioral harms that could arise from an inappropriate use of a particular approach.
For example, if the platform is in its early stages and does not have many customers, promotional (and especially brand-awareness) approaches can help increase platform adoption and gain a large customer base.
Conversely, if the platform already has many customers, profit-awareness (and particularly long-term value sustainability) approaches can boost corporate profitability.
Similarly, if many customers are leaving the platform, the business may employ utilitarian or promotional approaches to increase retention and build customer loyalty.
In addition, price-sensitivity methods may be useful in cases where price is an important driver in the user's final choice.
However, the company should consider to balance price-sensitivity appropriately with profit-awareness to avoid recommending unprofitable items.

Much work has been done on RSs since their first formulation and many different approaches were proposed over the years.
However, even if these systems were originally built to support business decisions, not much literature has yet focused on designing ECRSs to directly optimize organizational profitability.
This review shall help academic scholars and industry partners to navigate the existing literature and understand the state-of-the-art.
We hope this work will serve as a valuable starting point to foster future research and shift academic efforts towards more impactful RSs research that matters \cite{jannach_recommendations_2016, jannach_escaping_2020}.

\section{Acknowledgments}
This work was partially funded by estilos srl.

\bibliographystyle{cas-model2-names}

%\bibliography{references}

\end{document}